\documentclass[12pt]{article} 

\usepackage[utf8]{inputenc} 

\usepackage{geometry} 
\usepackage{graphicx} 

\usepackage{booktabs} 
\usepackage{array} 
\usepackage{paralist} 
\usepackage{verbatim} 
\usepackage{subfig} 

\usepackage{fancyhdr} 
\usepackage{sectsty}
\allsectionsfont{\sffamily\mdseries\upshape} 
\usepackage{lineno,hyperref}
\modulolinenumbers[5]
\usepackage{amsmath} 
 \usepackage[numbers,sort&compress]{natbib} 
\usepackage{amssymb}
 \usepackage{indentfirst} 
\usepackage{CJK}
\usepackage{makecell}
\usepackage{multirow}
\usepackage{cases}
 \usepackage{caption} 
\usepackage{pdflscape}
\usepackage[mathscr]{eucal}
\newcommand{\tabincell}[2]{\begin{tabular}{@{}#1@{}}#2\end{tabular}}

\date{} 
\title{Convective meta-thermal concentration for ultrahigh efficient Stirling engine with waste heat and cold utilization}
\author{Xinchen Zhou\textsuperscript{a}, Xiang Xu\textsuperscript{b}, Xiaoping Ouyang\textsuperscript{c,}*, Jiping Huang\textsuperscript{a,}*}

\begin{document}
\maketitle

\noindent {{\textsuperscript{a}Department of Physics, State Key Laboratory of Surface Physics, and Key Laboratory of Micro and Nano Photonic Structures (MOE), Fudan University, Shanghai 200433, China; \textsuperscript{b}Key Lab of Smart Prevention and Mitigation of Civil Engineering Disasters of the Ministry of Industry and Information Technology and Key Lab of Structures Dynamic Behavior and Control of the Ministry of Education, Harbin Institute of Technology, Harbin 150090, China; \textsuperscript{c}School of Materials Science and Engineering, Xiangtan University, Xiangtan 411105, China.}}

\  

\noindent* To whom correspondence should be addressed; E-mail: oyxp2003@aliyun.com, jphuang@fudan.edu.cn

\

\begin{abstract}
The Stirling engine, which possesses external combustion characteristics, a simple structure, and high theoretical thermal efficiency, has excellent potential for utilizing finite waste heat and cold resources. However, practical applications of this technology suffered from thermal inefficiency due to the discontinuity and instability of waste resources. Despite advances in energy storage technology, temperature variations in the heat-exchanging fluids at the hot and cold ends of the Stirling engine remained significant obstacles. In this work, convective meta-thermal concentration (CMTC) was introduced between the heating (cooling) fluids and the hot (cold) end of the Stirling engine, employing alternating isotropic materials with high and low thermal conductivities. It was demonstrated that CMTC effectively enhanced the temperature difference between the hot and cold ends, leading to a remarkable improvement in Stirling engine efficiency. Particularly, when the Stirling engine efficiency tended to zero due to the limited availability of waste heat and cold resources, CMTC overcame this limitation, surpassing existing optimization technology. Further analysis under various operating conditions showed that CMTC achieved a significant thermal efficiency improvement of up to 1460\%. This work expanded the application of thermal metamaterials to heat engine systems, offering an exciting avenue for sustainable energy utilization.
\end{abstract}

\noindent Keywords: Convective meta-thermal concentration, Stirling engine, ultrahigh efficient, waste heat and cold utilization.

\section{Introduction}

Actively harnessing waste heat and cold generated during energy-intensive processes, such as industrial production and liquefied natural gas (LNG), offers a practical and effective approach to enhancing energy efficiency and optimizing resource utilization \cite{JI2022,EMADI2020114384}. One approach involves utilizing a thermodynamic cycle to convert waste heat and cold into usable power. For instance, in the case of an LNG-fueled ship, Han et al. \cite{HAN2019561} developed an integrated power generation system that employed triple organic Rankine cycles (ORC) to recover waste heat and cold from the main engine and LNG fuel, respectively. This innovative setup successfully fulfilled the requirements of increased net output power, improved process temperature matching, and cost-effectiveness, among others. Further exergy analysis suggests that replacing the existing ORC with a Stirling engine could yield additional enhancements to the system's performance.

The Stirling engine possesses several main advantages, including inherent external combustion characteristics, a simple structure, and high theoretical thermal efficiency, making it suitable for any form of heat and cold sources. Previous research has demonstrated that solar energy, biomass energy, industrial waste heat, and more can all be used as heat sources \cite{BABAZADEH2023108290,ALNIMR2023894, BARTELA2018601,YUN2021106508}, while water, air, LNG cold energy, low-temperature waste heat, and others can be used as cold sources \cite{HAN2019561,ALNIMR2023894,HOU2018389}. By transferring heat and cold energy to both ends of the Stirling engine through heat exchangers, commonly referred to as the heater and cooler, continuous external work can be achieved. Therefore, the Stirling engine has great potential in the utilization of waste heat and cold.

Wang et al. \cite{WANG2017280} utilized a thermoacoustic Stirling electric generator to harness LNG cold energy and waste heat. The numerical study conducted by DeltaEC provides constructive guidelines for designing such thermoacoustic Stirling engine electric generators. Although the Stirling engine can theoretically achieve the same thermal efficiency as the Carnot cycle under ideal conditions, actual testing or analysis results often fall short of this requirement \cite{Sowale2019, UDEH2021}. The inefficiency of the Stirling engine can be attributed to three main factors. Firstly, thermal and hydraulic losses in the Stirling engine components contribute to inefficiency. Secondly, the heat transfer performance between the heat/cold source and the two ends of the Stirling engine affects efficiency. Lastly, the temperature variation of the heat and cold sources impacts performance. To address these issues, various research studies have been conducted.

Jiang et al. \cite{JIANG2023120242} designed a novel multi-stage Stirling generator with a multiple-bypass configuration, which outperformed traditional one-stage regenerators by improving electrical power by 29.3\%. Additionally, the researchers analyzed the influence of heating conditions and area ratios on the system's performance. The results showed that a temperature difference ratio of 0.75, an initial temperature of 1050 K, and a cone-shaped Stirling engine design were beneficial for system performance. Munir et al. \cite{MUNIR2020100664} proposed a new mini-channel heater with a variable cross-section and analyzed its performance under oscillating flow. The heat transfer rate was improved by a maximum of 27.6\% by increasing the dimensionless fluid displacement. Lai et al. \cite{LAI2016218} utilized Hitec salt doped with Sn\textsubscript{x}Zn\textsubscript{1-x}/SiO\textsubscript{x} core-shell alloy particles, which exhibited an endothermic plateau within the range of 643.15 K to 680.15 K. By applying these materials to the Stirling engine system as a heat source, the energy output improved by 21\% compared to using pure salt. Abuelyamen et al. \cite{ABUELYAMEN201753} conducted a parametric study on a $\mathrm{\beta}$-type Stirling engine using ANSYS Fluent 14.5. They examined the variable thermal properties of three different types of working fluids: air, helium, and hydrogen. By analyzing heat transfer amount, power output, and thermal efficiency, they concluded that the best engine performance was achieved when hydrogen was used as the working fluid. These research findings indicate that structural optimization, careful selection of operating conditions, materials development, and enhancement of thermal properties are beneficial for improving Stirling engine efficiency.

When considering the working conditions of Stirling engines with waste heat and cold utilization, two factors critically influence thermal efficiency: (1) The discontinuity and instability of resources \cite{LI2020}; (2) The finite heat capacity rate of the actual heating (cooling) fluids \cite{KIM2022,WHITE2020}. The first factor can be addressed with heat (cold) energy storage technology, which absorbs waste heat and cold when they are abundant and releases them when they are deficient [Fig. 1(a)] \cite{ZHANG2022E,ZRQ}. However, the second factor is often limited by the finite waste heat (cold) resources and the thermal properties of the actual heating (cooling) fluids \cite{LI2020,KIM2022,WHITE2020}. Due to the limitations of the actual heat-exchanging fluids' thermal properties, the temperature of these fluids can vary, which impedes the thermal efficiency of the Stirling engine based on Newton's cooling law.

Several studies have captured the temperature variation characteristics of heat and cold sources \cite{TLILI2012,AHMADI2014,AHMADI2015,PRAJAPATI2023127253}. Tlili \cite{TLILI2012} performed a finite-time thermodynamic analysis on an endoreversible Stirling engine, where the heat transfer between the working fluid and the heat source/sink was finite. The fluid temperature was approximately expressed by the logarithmic mean temperature. The results showed that the regenerator effectiveness and the heat capacity of the heat sink were crucial for the Stirling engine's performance. Ahmadi et al. \cite{AHMADI2014} considered the finite heat capacities of external reservoirs in a parametric demonstration of an irreversible Stirling cryogenic refrigerator cycle. They developed a multi-objective optimization method based on the Pareto optimal frontier, parallel to the single-objective optimization method. This method was then extended to the Stirling heat pump system \cite{AHMADI2015}, as well as recently proposed multi-objective ecological optimization of an irreversible Stirling cryogenic refrigerator cycle \cite {PRAJAPATI2023127253}. However, little research has focused on the intricate heat transfer performance between the hot (cold) end of the Stirling engine and the heat (cold) source, and the subsequent coupling effect on the engine's efficiency. In this context, it is crucial to enhance the heat transfer performance of the heater and cooler in Stirling engine systems. Traditional technologies, such as improving the heat transfer coefficient and increasing the heat transfer area \cite{MUNIR2020100664,SONG2015,GHEITH2015, LAAZAAR2022}, may be constrained in finite waste heat and cold utilization. The variable temperature characteristics of the heat-exchanging fluids increase the influence proportion of the heat transfer temperature difference. Therefore, developing a strategy to overcome this obstacle is crucial for improving Stirling engine efficiency in waste heat and cold utilization.

\begin{figure*}[h!]
\centering
\includegraphics[width=12cm]  {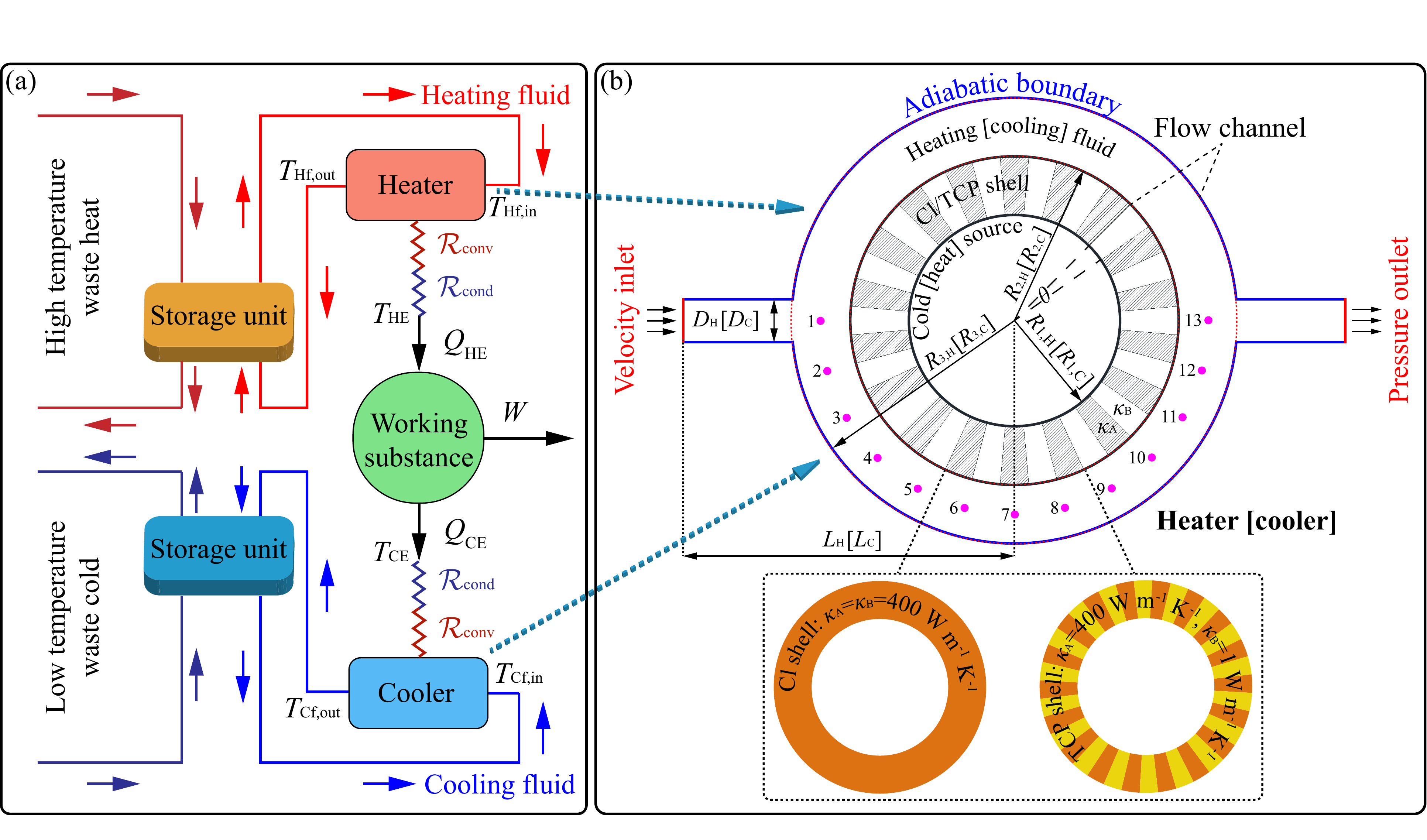}
\caption{\label{Fig. 1} Schematic diagram of (a) a Stirling engine system with waste heat and cold utilization and (b) the heater (cooler) under convective heating (cooling) modes.}
 \end{figure*}
 
Fortunately, the development of thermal metamaterials has provided a solution to the aforementioned problem \cite{HJP, XLJ, Zhang2023, YANG2021, JinPPNAS,zhu2015converting,xu2021geometric,xu2020transformation,xu2019thermal,yang2017full,dai2018transient,xu2018thermal,xu2022diffusive,xu2020transformation,
shen2016thermal,xu2020active,jin2020making,yang2019thermal}. This has led to the creation of a thermal device called a thermal concentrator, which is capable of significantly increasing the temperature gradient magnitude inside a shell compared to its surroundings \cite{YANG2021, ZPF, SXY, LI2016, YFB1, Narayana}. The thermal concentrator can be fabricated using two types of natural materials with high and low thermal conductivities arranged alternately, enabling various applications such as thermal energy harvesting, improvement of thermoelectric efficiency, and uniform heating \cite{HANTC2015, LiYM2022, HTC2}.

While these applications are primarily designed for conducting heat transfer, a similar thermal concentration function can be achieved under convective heat transfer. To illustrate this concept, Fig. 1(b) depicts two structures for the shells of the heater and cooler. One structure is made of a classical high thermal conductivity material (referred to as material A), called a Cl shell. The other structure is made by alternatively distributing materials with high and low thermal conductivities (materials A and B), resembling a thermal conductivity platter, termed a TCP shell. The thermal conductivities of materials A and B, denoted as $\kappa_\mathrm{A}$ and $\kappa_\mathrm{B}$, are set as 400 $\mathrm{W\  m^{-1}\  K^{-1}}$ and 1 $\mathrm{W\  m^{-1}\  K^{-1}}$, respectively, based on the thermal conductivities of copper and brick \cite{YSMTWQ}. It is important to note that this configuration is purely demonstrative, and the actual choice of materials can be tailored to specific requirements.

As shown in Fig. 2(a1,a2), constructing a TCP shell using alternating materials with high and low thermal conductivities enhances the heat transfer performance between the fluid outside the shell and the working substance (heat/cold source) inside the shell, compared to the Cl shell. The heat fluxes between the heating/cooling fluids and the working substance inside the shell are significantly regulated, transitioning from tangential flow to radial flow. These phenomena closely resemble the behavior of thermal concentrators under conduction heat transfer. Therefore, the passive temperature regulation achieved by the TCP shell is considered convective meta-thermal concentration (CMTC). Fig. 2(b1,b2) provides a more detailed representation of CMTC. It demonstrates that the average temperature of the substance inside the shell can be increased (or decreased) by raising (or lowering) the average temperatures of the heating (or cooling) fluids. This suggests that CMTC has the potential to improve the efficiency of Stirling engines without requiring additional energy input, enhancements in thermal properties, or an increase in heat transfer areas. As a result, CMTC offers an attractive option for effectively utilizing finite waste heat and cold resources.

 \begin{figure*}[htpb]
\centering
\includegraphics[width=12 cm]  {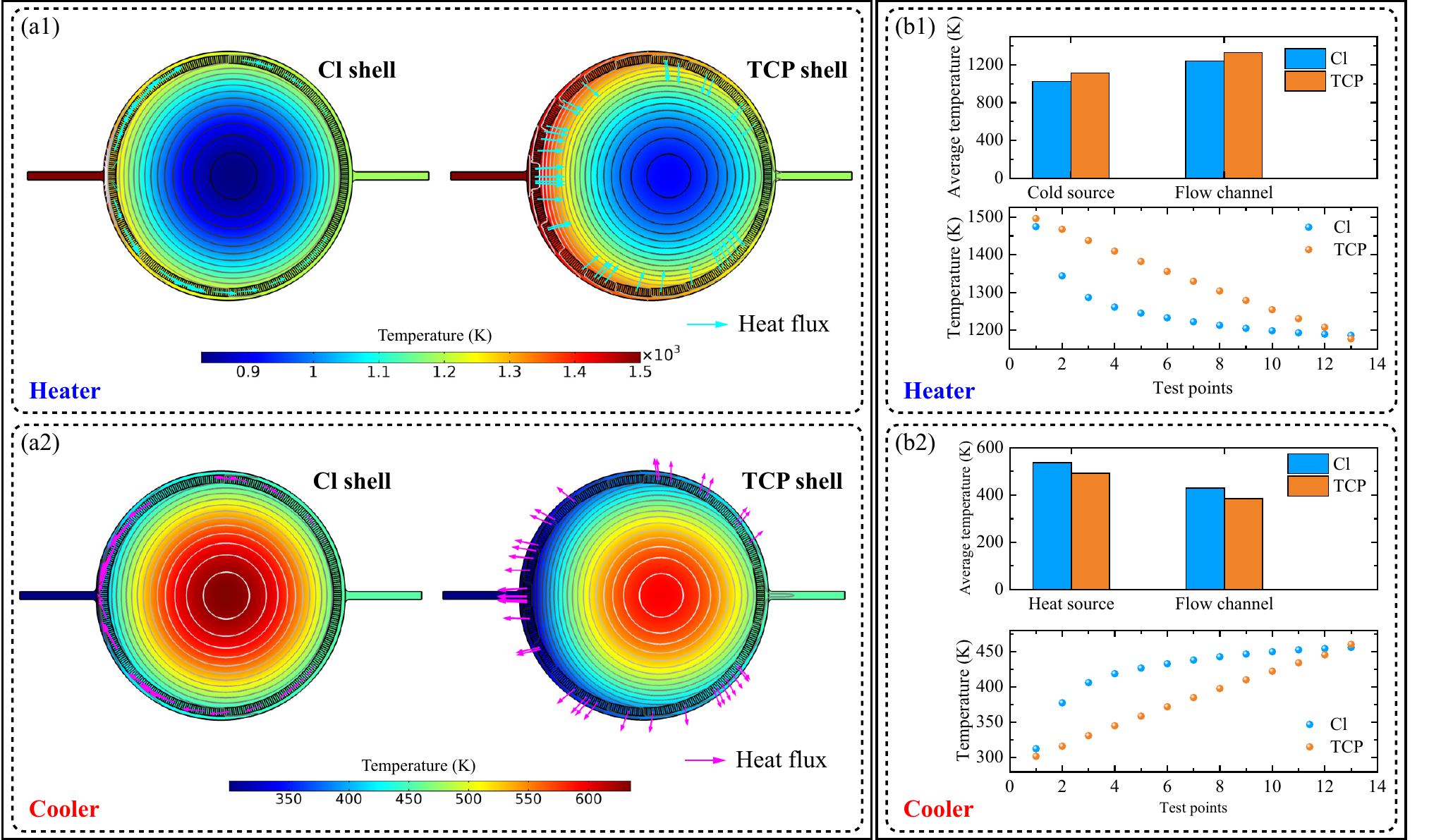}
\caption{\label{Fig. 2} (a1) [(a2)] Temperature distribution of the heater [cooler] with the Cl/TCP shell under convective heating [cooling] mode. (b1) [(b2)] Average temperature of the internal cold [heat] source \& the heating [cooling] fluid in the flow channel and the temperatures of characteristic test points [Fig. 1(b)]. The operating conditions are shown in Table 1. }
 \end{figure*}

The motivation of this study is to explore the potential of CMTC in enhancing the efficiency and power output of Stirling engines, particularly when operating with limited waste heat and cold resources. The aim is to devise effective strategies for improving energy utilization efficiency. To achieve this, a comprehensive framework is developed for a Stirling engine that incorporates waste heat and cold utilization by integrating the heating and cooling processes within the engine system. In contrast to existing technologies, this study focuses on enhancing the heat transfer performance of the heater and cooler through CMTC, thereby overcoming the challenges posed by temperature variations in the heat-exchanging fluids. The research demonstrates that CMTC has the potential to significantly improve Stirling engine efficiency, as evidenced by the analysis of various engine parameters under different operating conditions. The successful application of thermal concentration in a conduction heat transfer system using two alternately arranged isotropic materials, as previously reported in studies such as \cite{Narayana,HTC2,SXY,HANTC2015}, suggests that CMTC based on similar structures can be realized in the Stirling engine system. By expanding the application of thermal metamaterials to heat engine systems, this work presents a new avenue for designing highly efficient Stirling engines that effectively utilize waste heat and cold resources. The findings and results offer valuable insights and pave the way for the development of ultrahigh efficient Stirling engines with enhanced waste heat and cold utilization.

\section{Methodology}
\subsection {Model construction}
Fig.1(a) illustrates a Stirling engine system designed for the utilization of waste heat and cold. In the figure, $\mathcal{R}_\mathrm{cond}$ and $\mathcal{R}_\mathrm{conv}$ represent the conductive and convective thermal resistances between the heating (cooling) fluid and the working substance on the hot (cold) end, respectively. The inlet and outlet temperatures of the heating fluid in the heater are denoted as $T_\mathrm{Hf,in}$ and $T_\mathrm{Hf,out}$, while $T_\mathrm{Cf,in}$ and $T_\mathrm{Cf,out}$ represent the inlet and outlet temperatures of the cooling fluid in the cooler. The temperatures of the working substance on the hot and cold ends are denoted as $T_\mathrm{HE}$ and $T_\mathrm{CE}$, respectively. The Stirling engine receives input heat $Q_\mathrm{HE}$ and produces output heat $Q_\mathrm{CE}$, and the engine power is represented as $W$. To address the challenges of discontinuity and instability in waste heat (cold) resources, a heat (cold) storage unit is incorporated into the system. The working fluid passes through the heat (cold) storage unit to obtain (release) heat, which then forms the heating (cooling) fluid. Subsequently, the heating fluid flows through the heater of the Stirling engine, transferring heat to the working substance on the hot end. This heat transfer causes the working substance to expand and generate work output. Next, the heat from the working substance on the cold end is transferred to the cooling fluid in the cooler, resulting in a compression process. The Stirling engine performs these processes cyclically, enabling continuous conversion of heat to work.

Fig. 3 presents a Stirling engine model that incorporates detailed heating and cooling processes in the heater and cooler. The design of the shells for the heater and cooler can be approached in two ways. One approach involves constructing the shells using the Cl shell, while the other uses the TCP shell. The TCP shell represents the case of a Stirling engine with CMTC, whereas the Cl shell denotes a Stirling engine without CMTC. Specifically, Fig. 3(a) represents a Stirling engine model without CMTC, where the working substance in the Stirling engine is simplified as a green block. The input [output] heat flux $\dot Q_\mathrm{HE}^u$ [$\dot Q_\mathrm{CE}^u$] ($u=\mathrm{Cl}$ and $\mathrm{TCP}$) corresponds to the internal cold [heat] source density of the convective heating [cooling] modes for the heater [cooler] ($\dot Q_\mathrm{Ch}^u$ [$\dot Q_\mathrm{Hh}^u$]). The heater and cooler play a vital role in the convection heat transfer performance between the working substance and the waste heat and cold resources, as depicted in Fig. 3(b). To enhance the efficiency of heat exchange, the heater and cooler are coupled with CMTC. This coupling allows for increased and decreased temperatures of the Stirling engine's working substance inside the heater and cooler, respectively, as shown in Fig. 3(c). Fig. 3(d) illustrates the incorporation of this operation into a Stirling engine, representing a Stirling engine with CMTC. The heat transfer enhancement described in Fig. 3(c) compared to Fig. 3(b) can increase the output power density of a Stirling engine from Fig. 3(a) to Fig. 3(d). By analyzing the engine parameters of Stirling engines with and without CMTC, the effect of CMTC on improving Stirling engine efficiency can be revealed. The following analysis provides a detailed model assessment.

 \begin{figure}[h!]
\centering
\includegraphics[width=12cm]  {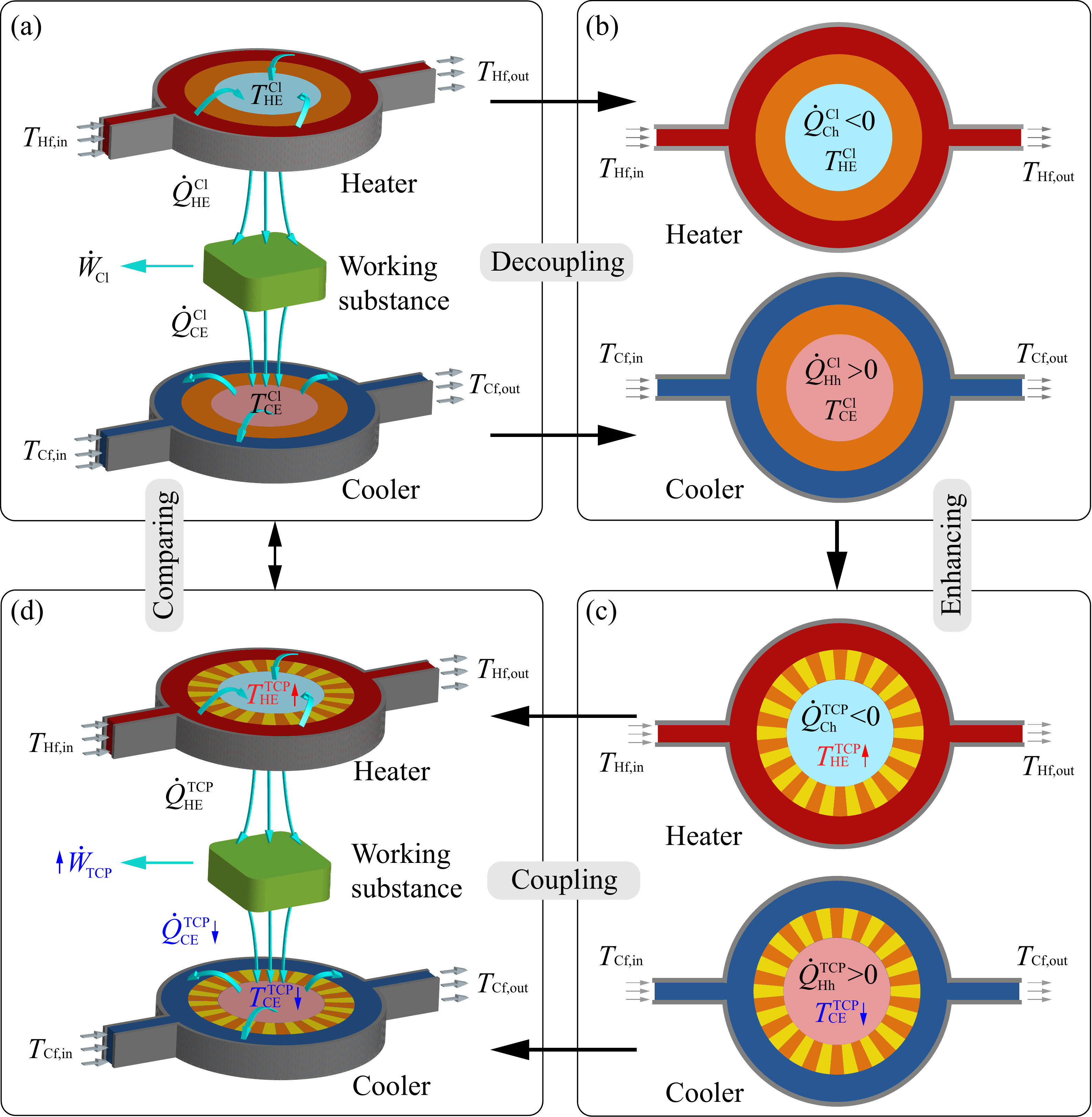}
\caption{\label{Fig. 3} Schematic diagram of the Stirling engine system with and without CMTC.  }
 \end{figure}
When determining the operating conditions of the heating (cooling) fluid, there are relationships between the average temperature of the working substance ($T_\mathrm{HE}^u$ [$T_\mathrm{CE}^u$]) inside the heater [cooler] and the input [output] heat ($Q_\mathrm{HE}^u$ [$Q_\mathrm{CE}^u$])
\begin{subequations}
\begin{align}
T_{\mathrm{HE}}^{u}=\mathscr{F}_1^u\left( Q_{\mathrm{HE}}^{u} \right),\\
T_{\mathrm{CE}}^{u}=\mathscr{F}_2^u\left( Q_{\mathrm{CE}}^{u} \right).
\end {align}
\end{subequations}
Here, the subscript ``$u$'' represents the types of shell. $u=\mathrm{Cl}$ and $u=\mathrm{TCP}$ denote the heater and cooler are equipped with Cl and TCP shells, respectively.  $\mathscr{F}_i^{u}\  \left(i=1,2,3\cdots\right)$ represents a mapping relation. It relies on the operating conditions of the heating (cooling) process. 

According to the geometries of the heater and cooler shown in Fig. 1(b), the relationships between input [output] heat ($Q_\mathrm{HE}^u$ [$Q_\mathrm{CE}^u$]) and input [output] heat flux ($\dot Q_\mathrm{HE}^u$ [$\dot Q_\mathrm{CE}^u$]) are
\begin{subequations}
\begin{align}
	Q_{\mathrm{HE}}^{u}=S_\mathrm{H}d_\mathrm{H}\dot{Q}_{\mathrm{HE}}^{u},\\
	Q_{\mathrm{CE}}^{u}=S_\mathrm{C}d_\mathrm{C}\dot{Q}_{\mathrm{CE}}^{u},
\end{align}
\end{subequations}
where $S_\mathrm{H}$ [$S_\mathrm{C}$] is the area of the internal cold [heat] source of the heater [cooler], $d_\mathrm{H}$ [$d_\mathrm{C}$] is the thickness of the heater [cooler]. Then, the $Q_\mathrm{HE}^u$ and $Q_\mathrm{CE}^u$ in Eq. (1a) and Eq. (1b) were replaced with $\dot Q_\mathrm{HE}^u$ and $\dot Q_\mathrm{CE}^u$, respectively,
\begin{subequations}
\begin{align}
T_{\mathrm{HE}}^{u}=\mathscr{F}_3^{u}\left( \dot{Q}_{\mathrm{HE}}^{u} \right),\\
T_{\mathrm{CE}}^{u}=\mathscr{F}_4^{u}\left( \dot{Q}_{\mathrm{CE}}^{u} \right).
\end{align}
\end{subequations}
Eq. (3a) and Eq. (3b) are referred to as the heating and cooling correlations of the heater and cooler, respectively. 

Then, the heating and cooling correlations were coupled with the Stirling engine model. Before that, it is necessary to explain the difference between the Stirling engine and the ideal Carnot engine. Fig. 4(a) illustrates an ideal Carnot cycle, where the process $1\rightarrow2$ represents the isothermal endothermic process, $2\rightarrow 3$ is the adiabatic expansion process, $3\rightarrow 4$ denotes the isothermal exothermic process, and $4\rightarrow1$ is the adiabatic compression process. The ideal Carnot cycle efficiency is
\begin {equation}
\eta_\mathrm{Carnot}=1-\frac{T_\mathrm{Cf,in}}{T_\mathrm{Hf,in}}.
\end {equation}
For comparison, Fig. 4(b) presents an ideal Stirling cycle, where the process $1\rightarrow2$ represents the isothermal expansion process, $2\rightarrow3$ is the isovolumic exothermic process, $3\rightarrow4$ denotes the isothermal compression process, and $4\rightarrow1$ is the isovolumic endothermic process. When the regenerator of the Stirling engine is ideal, the Stirling engine efficiency is equivalent to that of an ideal Carnot engine. However, in the real world, the heat transfer temperature difference needs to be considered. Therefore, the hot (cold) end temperature of the Stirling engine is lower (higher) than that of the ideal Carnot engine. The average temperature of the working substance at the hot (cold) end is considered the temperature of the hot (cold) end of the Stirling engine.  Specifically, for Fig. 4(c) [Fig 4(d)], $T_\mathrm{HE}^\mathrm{Cl}$ and $T_\mathrm{CE}^\mathrm{Cl}$ [$T_\mathrm{HE}^\mathrm{TCP}$ and $T_\mathrm{CE}^\mathrm{TCP}$] were treated as the temperature of the $1\rightarrow2$ and $3\rightarrow4$ processes, respectively. Assuming the Stirling engine undergoes an internal reversible cycle, its thermal efficiency [Fig. 4 (c) and (d)] reads
\begin {equation}
\eta_u=1-\frac{T_\mathrm{CE}^u\left(\dot Q_\mathrm{CE}^u\right)}{T_\mathrm{HE}^u\left(\dot Q_\mathrm{HE}^u\right)}.
\end {equation}

 \begin{figure}[h!]
\centering
\includegraphics[width=12cm]  {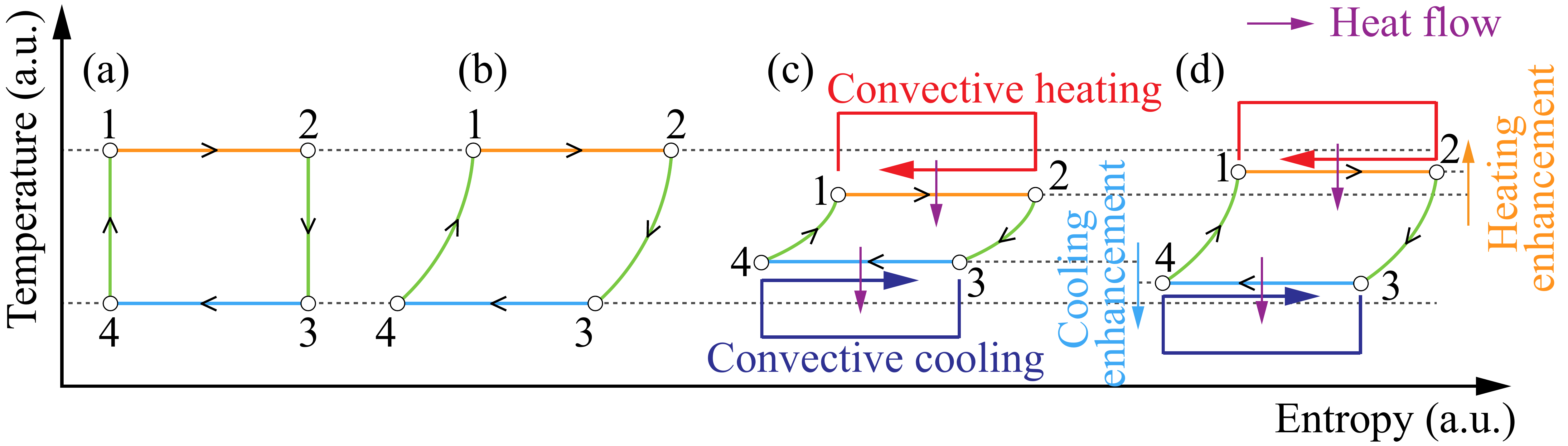}
\caption{\label{Fig. 4} Schematic of the temperature-entropy diagram for various thermodynamic cycles.}
 \end{figure} 
\noindent Note that, in Eq. (5), the independent variables representing the temperatures at the hot and cold ends, $\dot Q_\mathrm{HE}^u$ and $\dot Q_\mathrm{CE}^u$, correspond to the heat flux absorbed and released by the working substance of the Stirling engine, respectively. In a heat engine system, $\dot Q_\mathrm{CE}^u$ is related to $\dot Q_\mathrm{HE}^u$. Under the reversible condition, there exists
\begin {equation}
\frac{T_{\mathrm{CE}}^{u}}{T_{\mathrm{HE}}^{u}}=\frac{Q_{\mathrm{CE}}^{u}}{Q_{\mathrm{HE}}^{u}}=\frac{S_\mathrm{C}d_\mathrm{C}\dot{Q}_{\mathrm{CE}}^{u}}{S_\mathrm{H}d_\mathrm{H}\dot{Q}_{\mathrm{HE}}^{u}}.
\end {equation}
According to Eqs. (3) and (6), $\dot Q_{\mathrm{CE}}^u$ in Eq. (5) can be replaced by a function related to $\dot Q_{\mathrm{HE}}^u$
\begin {equation}
\dot Q_\mathrm{CE}^u=\mathscr{F}_5^u(\dot Q_\mathrm{HE}^u).
\end {equation}
Thus, the relationship between Stirling engine efficiency ($\eta_u$) and input heat flux ($\dot Q_\mathrm{HE}^u$) can be expressed as
\begin {equation}
\eta_u=1-\frac{Q_\mathrm{CE}^u}{Q_\mathrm{HE}^u}=1-\frac{S_\mathrm{C}d_\mathrm{C}\mathscr{F}_5^u(\dot Q_\mathrm{HE}^u)}{S_\mathrm{H}d_\mathrm{H}\dot Q_\mathrm{HE}^u}=\mathscr{F}_6^u(\dot Q_\mathrm{HE}^u).
\end {equation}
By now, a Stirling engine model incorporating detailed heating and cooling processes, with waste heat and cold utilization, has been constructed.

\subsection{Methods for parameters visualization}
As mentioned above, the heating and cooling correlations [Eq. (3)] are the basis for visualizing the Stirling engine parameters. This goal was realized through finite element simulations (COMSOL Multiphysics). The specific methods are as follows.

\subsubsection{Geometry,  grid division, and boundary conditions}
The heat transfer processes in the heater and cooler are explained in detail in Sec. 2.1. For simulations, the corresponding geometric model rendered by the grid for the heater/cooler is shown in Fig. 5, with the following structural parameters: $D_\mathrm{H}$ [$D_\mathrm{C}$]=0.2 $\mathrm{mm}$, $L_\mathrm{H}$ [$L_\mathrm{C}$]=5 $\mathrm{mm}$, $R_\mathrm{1,H}$ [$R_\mathrm{1,C}$]=2.8 $\mathrm{mm}$, $R_\mathrm{2,H}$ [$R_\mathrm{2,C}$]=3.0 $\mathrm{mm}$, $R_\mathrm{3,H}$ [$R_\mathrm{3,C}$]=3.1 $\mathrm{mm}$, and $\theta=1^\circ$. In the heater (cooler), the internal of the shell corresponds to the working substance of the hot (cold) end in the Stirling engine. Therefore, it can be treated as an internal cold (heat) source from the perspective of the heat transfer model. The detailed description for each zone of the heat transfer model and its boundary conditions are marked. The left and right boundaries are the velocity inlet and pressure outlet, respectively. The upper and lower boundaries are adiabatic. The detailed descriptions for $T_\mathrm{in}$ and $p_\mathrm{out}$ are shown in Table 1.
\begin{figure}[h!]
\centering
\includegraphics[width=8cm]  {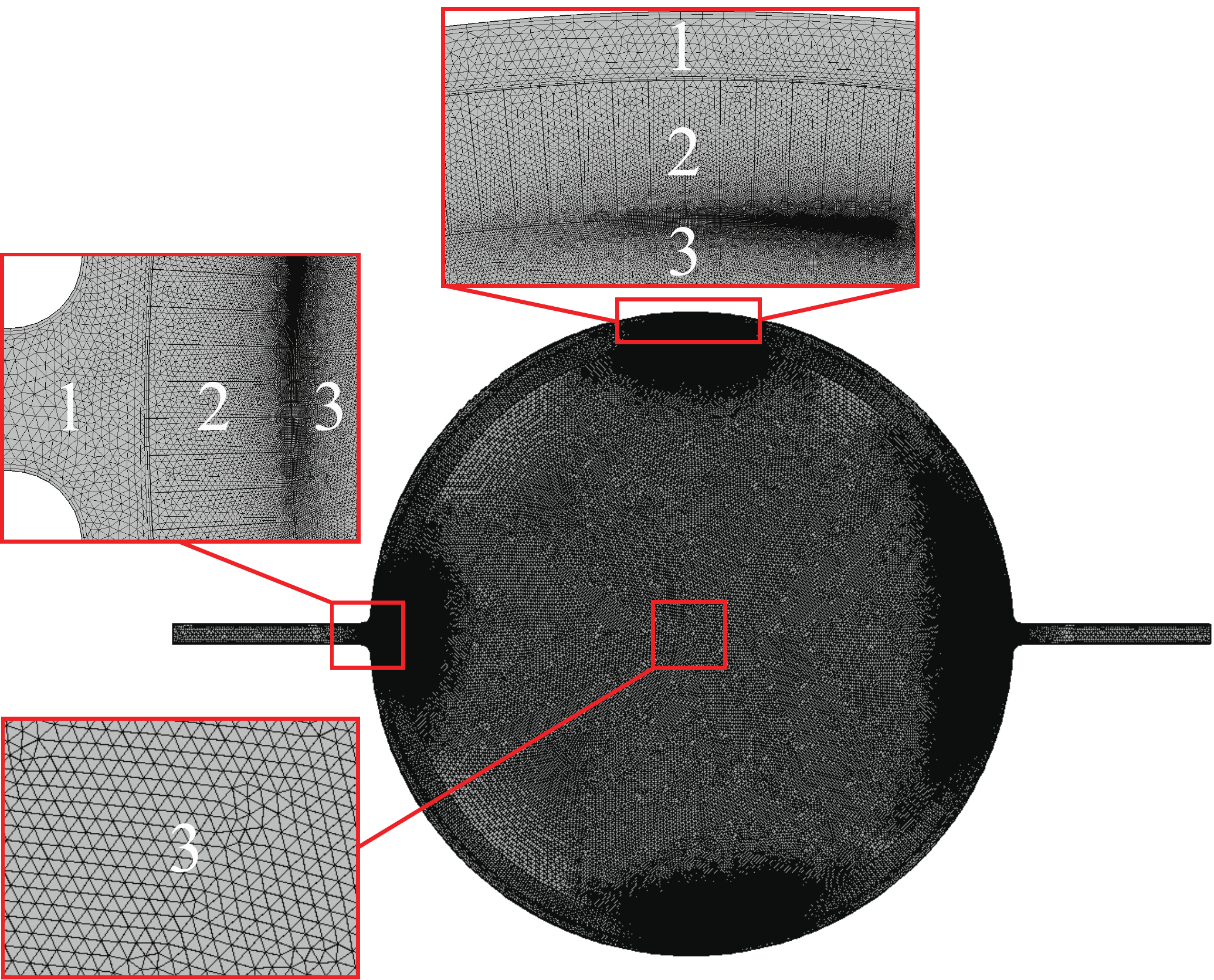}
\caption{\label{Fig. 5} Schematic diagram of grid division for the heater/cooler.}
 \end{figure}
\subsubsection{Governing equations}
To focus on the CMTC effect on Stirling engine efficiency improvement, the following simplifications were made:

(1) The working fluids in the heater and cooler were considered to be incompressible Newtonian fluids.

(2) All thermal properties were assumed to be independent of temperature.

(3) The system components were assumed to undergo no phase transition under all operating conditions.

(4) The heat transfer processes were assumed to be in a steady state.

Combining Fig. 5 with Fig. 1(b), it can be observed that Region 1 is filled with heating/cooling fluids, Region 2 is a solid shell, and Region 3 is an internal heat/cold source. The governing equations for each region are listed as follows. 

(1) For Region 1, there exists a forced convection heat transfer, which can be described as
\begin {equation}
\left\{\begin{aligned}
&\nabla \cdot \boldsymbol v =0\\
&\rho_\mathrm{f}(\boldsymbol v \cdot \nabla )\boldsymbol v =\boldsymbol F-\nabla p+\mu \nabla^2\boldsymbol v \\
&\rho_\mathrm{f}c_\mathrm{f}(\boldsymbol v \cdot \nabla T_\mathrm{f})=\kappa_\mathrm{f}\nabla^2T_\mathrm{f}+\dot \phi_\mathrm{f}
\end {aligned}
\right.,
\end {equation}
where $\boldsymbol v$, $\rho_\mathrm{f}$, $\boldsymbol F$, $p$, $\mu$, $\kappa_\mathrm{f}$,  $c_\mathrm{f}$, and $T_\mathrm{f}$ are the velocity, density, body force,  pressure, dynamic viscosity, thermal conductivity, specific heat capacity, temperature of heat-exchanging fluids, respectively. $\dot \phi_\mathrm{f}$ denotes heat source power caused by viscous dissipation.

(2) For Region 2, there exists a conduction heat transfer without an internal heat/cold source, which is controlled by
\begin {equation}
\nabla^2 T_\mathrm{s}=0,
\end {equation}
where $T_\mathrm{s}$ is the temperature distribution of the shell.

(3) For Region 3, there exists a conduction heat transfer with an internal heat/cold source. The governing equation reads
\begin {equation}
\nabla^2 T_\mathrm{I}+\frac{\dot\phi_\mathrm{I}}{\kappa_\mathrm{I}}=0,
\end {equation}
where $T_\mathrm{I}$, $\kappa_\mathrm{I}$, and $\dot \phi_\mathrm{I}$ are the temperature distribution, thermal conductivity, and power density of internal heat/cold source, respectively.

\subsubsection {Operating conditions}
To study the effects of various factors on CMTC's ability to improve Stirling engine efficiency, different operating conditions were set to simulate practical applications. The details of these operating conditions are presented in Table 1.
\begin{landscape}
\begin{table}[htpb]\scriptsize
\centering
\tabcolsep=0.15cm
\caption{Various operating conditions. In all cases, $\kappa_\mathrm{Hh}$ [$\kappa_\mathrm{Ch}$]=1 $\mathrm{W\  m^{-1}\  K^{-1}}$, $\mu_\mathrm{Hf}$ [$\mu_\mathrm{Cf}$]=0.01 $\mathrm{Pa\  s}$, $\gamma_\mathrm{Hf}$ [$\gamma_\mathrm{Cf}$]=1, and $p_\mathrm{Hf,out}$ [$p_\mathrm{Cf,out}$]=1.01 $\times 10^5$ $\mathrm{Pa}$. For Fig. 2, $\dot Q_\mathrm{Ch}^u=-2\times 10^8\  \mathrm{{W}\  m^{-3}}$ and $\dot Q_\mathrm{Hh}^u=1\times 10^8\  \mathrm{{W}\  m^{-3}}$.}
\begin{tabular}{ccccccccc}
\hline
Case &  \tabincell{c}{$c_\mathrm{Hf}$ [$c_\mathrm{Cf}$] \\ $(\mathrm{J\  kg^{-1}\  K^{-1}})$} & \tabincell{c}{$\rho_\mathrm{Hf}$ [$\rho_\mathrm{Cf}$] \\ $(\mathrm{kg\  m^{-3}})$}&\tabincell{c}{$\kappa_\mathrm{Hf}$ [$\kappa_\mathrm{Cf}$]  \\ $(\mathrm{W\  m^{-1}\   K^{-1}})$}& \tabincell{c}{$v_\mathrm{Hf,in}$ [$v_\mathrm{Cf,in}$] \\ $(\mathrm{m\  s^{-1}})$}&\tabincell{c}{ $T_\mathrm{Hf,in}$ [$T_\mathrm{Cf,in}$]\\ $(\mathrm{K})$} \\ \hline
Fig. 2  &  2000  &  1000 &  0.5 & 0.04 & 1500 [300]\\
Fig. 7     & 2000  &  1000 & 0.5  &  0.04 & \tabincell{c}{1100 [1000], 1200 [900], 1300 [800],\\1400 [700], 1500 [600], 1600 [500],\\1700 [400], 1800 [300], 1900 [200], 2000 [100]} \\
Fig. 8      & 2000  &  1000 & 0.5  &  \tabincell{c}{0.01, 0.02, 0.03,\\0.04, 0.05, 0.06,\\0.07, 0.08, 0.09, 0.10}  &1800 [300]\\
Fig. 9     & 2000  & \tabincell{c}{200, 300, 400, 500,\\600, 700, 800, 900,\\1000, 1500, 2000, 3000, 4000} & 0.5  &0.05&1800 [300]\\
Fig. 10    & \tabincell{c}{600, 700, 800,\\900, 1000, 1500,\\2000, 2500, 3000, 3500}  &2000& 0.5  &0.05&1800 [300]   \\
Fig. 11  & 2000 &2000& \tabincell{c}{0.5, 1, 2,\\3, 4, 5,\\6, 7, 8, 9}   &0.05&1800 [300]   \\
Fig. 12& 500 &500& \tabincell{c}{0.5, 1, 2,\\3, 4, 5,\\6, 7, 8, 9}   &0.05&1800 [300]   \\ \hline
\end{tabular}
\end{table}
\end{landscape}
\subsubsection{Grid independence analysis}
Following the description provided above, finite element simulations could be conducted. The next step involves selecting a suitable grid division scheme. As shown in Table 2, three schemes with different mesh numbers have been generated.

\begin{table}[!h]\scriptsize
\caption{Outlet temperatures of the heater and cooler under different mesh numbers. }
\centering
\renewcommand\arraystretch{1.1}
\begin{tabular}{cccccc}
\hline
Schemes & Mesh number  & $T_\mathrm{Hf,out}^{\mathrm{Cl}}$ $(\mathrm{K})$ & $T_\mathrm{Hf,out}^{\mathrm{TCP}}$ $(\mathrm{K})$ & $T_\mathrm{Cf,out}^{\mathrm{Cl}}$ $(\mathrm{K})$ & $T_\mathrm{Cf,out}^{\mathrm{TCP}}$ $(\mathrm{K})$ \\ \hline
1    & 348968  & 1263.6 & 1263.8  & 535.04 & 535.03  \\
2    & 1395872 & 1265.4 & 1265.3  & 532.45 & 532.53  \\
3    & 5577920 & 1266.7 & 1264.6  & 531.40 & 531.09  \\ \hline
\end{tabular}
\end{table}

Consider Fig. 2 as a case study. The influence of mesh numbers on the outlet fluid temperatures of the heater and cooler with Cl and TCP shells ($T_\mathrm{Hf,out}^{\mathrm{Cl}}$, $T_\mathrm{Hf,out}^{\mathrm{TCP}}$, $T_\mathrm{Cf,out}^{\mathrm{Cl}}$, and $T_\mathrm{Cf,out}^{\mathrm{TCP}}$) under steady simulation was tested by resetting $\dot Q_\mathrm{HE}^u=-1.5\times10^8\ \mathrm{W\ m^{-3}}$ and $\dot Q_\mathrm{CE}^u=1.5\times10^8\ \mathrm{W\ m^{-3}}$, while keeping the remaining operating conditions unchanged. The results are shown in Table 2. The accuracy of these results can be examined by theoretical calculations. The theoretical results of $T_\mathrm{Hf,out}^{u}$ is
\begin {equation}
\begin{aligned}
T_\mathrm{Hf,out}^{u}&=\frac{\pi R_\mathrm{1,H}^2\dot Q_\mathrm{HE}^u}{c_\mathrm{Hf}\rho_\mathrm{Hf}v_\mathrm{Hf,in}D_\mathrm{H}}+T_\mathrm{Hf,in}\\
&=\frac{\pi\times(2.8/1000)^2\times (-1.5\times 10^8)}{2000\times1000\times0.04\times0.2/1000}+1500\\
&=1269.09\  \mathrm{K}.
\end{aligned}
\end {equation}
The theoretical results of $T_\mathrm{Cf,out}^{u}$ is
\begin {equation}
\begin {aligned}
T_\mathrm{Cf,out}^{u}&=\frac{\pi R_\mathrm{1,C}^2\dot Q_\mathrm{CE}^u}{c_\mathrm{Cf}\rho_\mathrm{Cf}v_\mathrm{Cf,in}D_\mathrm{C}}+T_\mathrm{Cf,in}\\
&=\frac{\pi\times(2.8/1000)^2\times 1.5\times 10^8}{2000\times1000\times0.04\times0.2/1000}+300\\
&=530.91\  \mathrm{K}.
\end {aligned}
\end {equation}
The relative errors between the simulation results $\left(N_\mathrm{sim}\right)$ with the theoretical results $\left(N_\mathrm{the}\right)$ were calculated by
\begin {equation}
E=\frac{N_\mathrm{sim}-N_\mathrm{the}}{N_\mathrm{the}}\times 100\%.
\end {equation}
The results are presented in Table 3, where the magnitudes of the errors are below 1\%. It can be observed that increasing the number of mesh divisions leads to higher accuracy in the simulations.

\begin{table}[!h]\scriptsize
\tabcolsep=0.13cm
\caption{Relative errors of outlet temperatures with different mesh numbers.}
\centering
\renewcommand\arraystretch{1.1}
\begin{tabular}{cccccc}
\hline
Schemes & Mesh number  & $E\left(T_\mathrm{Hf,out}^{\mathrm{Cl}}\right)$ (\%) & $E\left(T_\mathrm{Hf,out}^{\mathrm{TCP}}\right)$ (\%) & $E\left(T_\mathrm{Cf,out}^{\mathrm{Cl}}\right)$ (\%) & $E\left(T_\mathrm{Cf,out}^{\mathrm{TCP}}\right)$ (\%) \\ \hline
1    & 348968  & -0.43 & -0.42  & 0.78 & 0.78  \\
2    & 1395872 & -0.29 & -0.30  & 0.29 & 0.31  \\
3    & 5577920 & -0.19 & -0.35  & 0.09 & 0.03  \\ \hline
\end{tabular}
\end{table}

Furthermore, the influence of mesh numbers on the average temperatures of the working substance at the hot and cold ends with Cl and TCP shells ($T_\mathrm{HE}^{\mathrm{Cl}}$, $T_\mathrm{HE}^{\mathrm{TCP}}$, $T_\mathrm{CE}^{\mathrm{Cl}}$, and $T_\mathrm{CE}^{\mathrm{TCP}}$) was tested under steady-state simulation, as shown in Table 4. Using the simulation results from case 3 as references ($N_\mathrm{ref}$), the relative errors of the other results were calculated by
\begin {equation}
M=\frac{N_\mathrm{sim}-N_\mathrm{ref}}{N_\mathrm{ref}}\times 100\%.
\end {equation}
The results are presented in Table 5. Since all the simulation results are close, it can be concluded that the results are independent of the mesh numbers. Considering both precision and efficiency, scheme 2 was chosen to perform all the simulations in this study.

\begin{table}[!h]\scriptsize
\centering
\caption{Average temperatures of the working substance on the hot and cold ends under different mesh numbers.}
\renewcommand\arraystretch{1.1}
\begin{tabular}{cccccc}
\hline
Schemes & Mesh number  & $T_\mathrm{HE}^{\mathrm{Cl}}\  (\mathrm{K})$ & $T_\mathrm{HE}^{\mathrm{TCP}}\  (\mathrm{K})$ & $T_\mathrm{CE}^{\mathrm{Cl}}\  (\mathrm{K})$ & $T_\mathrm{CE}^{\mathrm{TCP}}\  (\mathrm{K})$ \\ \hline
1    & 348968  & 1142.0& 1210.1  & 657.08 & 589.75  \\
2    & 1395872 & 1144.0 & 1210.8  & 654.71 & 588.90  \\
3    & 5577920 & 1145.1 & 1211.0  & 653.69 & 588.39  \\ \hline
\end{tabular}
\end{table}

\begin{table}[!h]\scriptsize
\centering
\tabcolsep=0.15cm
\caption{Relative errors of the average temperatures of the working substance on the hot and cold ends under different mesh numbers.}
\renewcommand\arraystretch{1.1}
\begin{tabular}{cccccc}
\hline
Schemes & Mesh number  & $M\left(T_\mathrm{HE}^{\mathrm{Cl}}\right)\  (\%)$ & $M\left(T_\mathrm{HE}^{\mathrm{TCP}}\right)\  (\%)$ & $M\left(T_\mathrm{CE}^{\mathrm{Cl}}\right)\  (\%)$ & $M\left(T_\mathrm{CE}^{\mathrm{TCP}}\right)\  (\%)$ \\ \hline
1    & 348968  & -0.27& -0.07  & 0.52 & 0.23  \\
2    & 1395872 & -0.10 & -0.02  & 0.16 & 0.09  \\
3    & 5577920 & Ref. & Ref. & Ref. & Ref.  \\ \hline
\end{tabular}
\end{table}

\subsection{Key factors definition}
As described above, CMTC enhances the heating process by increasing $T_\mathrm{HE}^\mathrm{Cl}$ to $T_\mathrm{HE}^\mathrm{TCP}$, corresponding to the increase of $T_\mathrm{Hf}^\mathrm{Cl}$ to $T_\mathrm{Hf}^\mathrm{TCP}$, as shown in Fig. 2(b1). The magnitude of this enhancement can be quantitatively described using a heating enhancement factor
\begin {equation}
\alpha=\frac{T_\mathrm{HE}^\mathrm{TCP}}{T_\mathrm{HE}^\mathrm{Cl}}=\frac{\mathscr{F}_3^\mathrm{TCP}\left(\dot Q_\mathrm{HE}^\mathrm{TCP}\right)}{\mathscr{F}_3^\mathrm{Cl}\left(\dot Q_\mathrm{HE}^\mathrm{Cl}\right)}=\mathscr{F}_7^u\left(\dot Q_\mathrm{HE}^u\right),\  \dot Q_\mathrm{HE}^\mathrm{TCP}=\dot Q_\mathrm{HE}^\mathrm{Cl}=\dot Q_\mathrm{HE}^u,\  \alpha>1.
\end {equation}
Similarly, in the cooling process, CMTC decreases $T_\mathrm{CE}^\mathrm{Cl}$ to $T_\mathrm{CE}^\mathrm{TCP}$ by reducing $T_\mathrm{Cf}^\mathrm{Cl}$ to $T_\mathrm{Cf}^\mathrm{TCP}$, as shown in Fig. 2(b2). A cooling enhancement factor is defined as
\begin {equation}
\beta_1=\frac{T_\mathrm{CE}^\mathrm{TCP}}{T_\mathrm{CE}^\mathrm{Cl}}=\frac{\mathscr{F}_4^\mathrm{TCP}\left(\dot Q_\mathrm{CE}^\mathrm{TCP}\right)}{\mathscr{F}_4^\mathrm{Cl}\left(\dot Q_\mathrm{CE}^\mathrm{Cl}\right)}=\mathscr{F}_8^u\left(\dot Q_\mathrm{CE}^u\right),\  \dot Q_\mathrm{CE}^\mathrm{TCP}=\dot Q_\mathrm{CE}^\mathrm{Cl}=\dot Q_\mathrm{CE}^u,\  \beta_1<1.
\end {equation}
 For ease of presentation, in Eq. (17), the independent variable $\dot Q_\mathrm{CE}^u$ was replaced with $\dot Q_\mathrm{HE}^u$. Substituting Eq. (7) into Eq. (17), it can be rewritten as
\begin {equation}
\beta_2=\frac{\mathscr{F}_4^\mathrm{TCP}\left[\mathscr{F}_5^{\mathrm{TCP}}\left(\dot Q_\mathrm{HE}^\mathrm{TCP}\right)\right]}{\mathscr{F}_4^\mathrm{Cl}\left[\mathscr{F}_5^{\mathrm{Cl}}\left(\dot Q_\mathrm{HE}^\mathrm{Cl}\right)\right]}=\mathscr{F}_9^u\left(\dot Q_\mathrm{HE}^u\right),\  \dot Q_\mathrm{HE}^\mathrm{TCP}=\dot Q_\mathrm{HE}^\mathrm{Cl}=\dot Q_\mathrm{HE}^u,\  \beta_2<1. 
\end {equation}
To quantitatively assess the effect of CMTC on improving Stirling engine efficiency, the improvement magnitude can be characterized by
\begin {equation}
\zeta=\frac{\eta_\mathrm{TCP}-\eta_\mathrm{Cl}}{\eta_\mathrm{Cl}}=\frac{T_\mathrm{CE}^\mathrm{Cl}/T_\mathrm{HE}^\mathrm{Cl}-T_\mathrm{CE}^\mathrm{TCP}/T_\mathrm{HE}^\mathrm{TCP}}{1-T_\mathrm{CE}^\mathrm{Cl}/T_\mathrm{HE}^\mathrm{Cl}}.
\end {equation}

Furthermore, the following factors were considered to explore the core influencing factors on $\zeta$.

(1) Inlet temperature difference
\begin{equation}
\Delta T_\mathrm{in}=T_\mathrm{Hf,in}-T_\mathrm{Cf,in}.
\end {equation}

(2) Heat capacity rate
\begin{equation}
\dot C_\varepsilon= c_\varepsilon \rho_\varepsilon v_\varepsilon D_\varepsilon d_\omega,
\end {equation}
where $c_\varepsilon$, $\rho_\varepsilon$, $v_\varepsilon$, and $D_\varepsilon$ are the specific heat capacity, density, average velocity, and inlet (outlet) width of the heating and cooling fluids, respectively. Here, $\varepsilon$ represents the heating and cooling fluids, with $\varepsilon=\mathrm{Hf}$ and $\varepsilon=\mathrm{Cf}$ denoting the heating and cooling fluids, respectively. $d_\omega$ is the thickness of the heater (cooler), and $\omega$ represents the heater and cooler, with $\omega=\mathrm{H}$ and $\omega=\mathrm{C}$ corresponding to the heater and cooler, respectively.

(3) Thermal diffusivity
\begin{equation}
\mathscr{A_\varepsilon}=\frac{\kappa_\varepsilon}{\rho_\varepsilon c_\varepsilon},
\end{equation} 
where $\kappa_\varepsilon$ is the thermal conductivity of the heating and cooling fluids. 

\section{Results and discussion}
\subsection{Model verification}
The accuracy of heat transfer processes in the heater and cooler has been verified in Sec. 2.2.4. This section focuses on evaluating the rationality of the general Stirling engine model with waste heat and cold utilization. The objective is to examine how various Stirling engine parameters change with respect to the input heat flux ($\dot Q_\mathrm{HE}^u$), ensuring consistency with existing results and capturing the detailed heat transfer characteristics in the heater and cooler.

Taking Fig. 2 as a case study, the simulation allows for determining the corresponding heating [cooling] correlation by adjusting $\dot Q_\mathrm{HE}^u$ [$\dot Q_\mathrm{CE}^u$] while keeping the operating conditions unchanged. This can be observed in Fig. 6(a,b). Notably, under the same $\dot Q^u_\mathrm{HE}$, the introduction of CMTC leads to higher [lower] hot [cold] end temperatures ($T_\mathrm{HE}^\mathrm{TCP}$ [$T_\mathrm{CE}^\mathrm{TCP}$]) compared to the absence of CMTC ($T_\mathrm{HE}^\mathrm{Cl}$ [$T_\mathrm{CE}^\mathrm{Cl}$]). As a result, the engine power density with CMTC ($\dot W_\mathrm{TCP}$) surpasses that without CMTC ($\dot W_\mathrm{Cl}$), resulting in an improved thermal efficiency from $\eta_\mathrm{Cl}$ to $\eta_\mathrm{TCP}$, as depicted in Fig. 6(c). Therefore, the Stirling engine model quantitatively reveals the impact of CMTC on Stirling engine efficiency.

\begin{figure}[h!]
\centering
\includegraphics[width=12.5cm]  {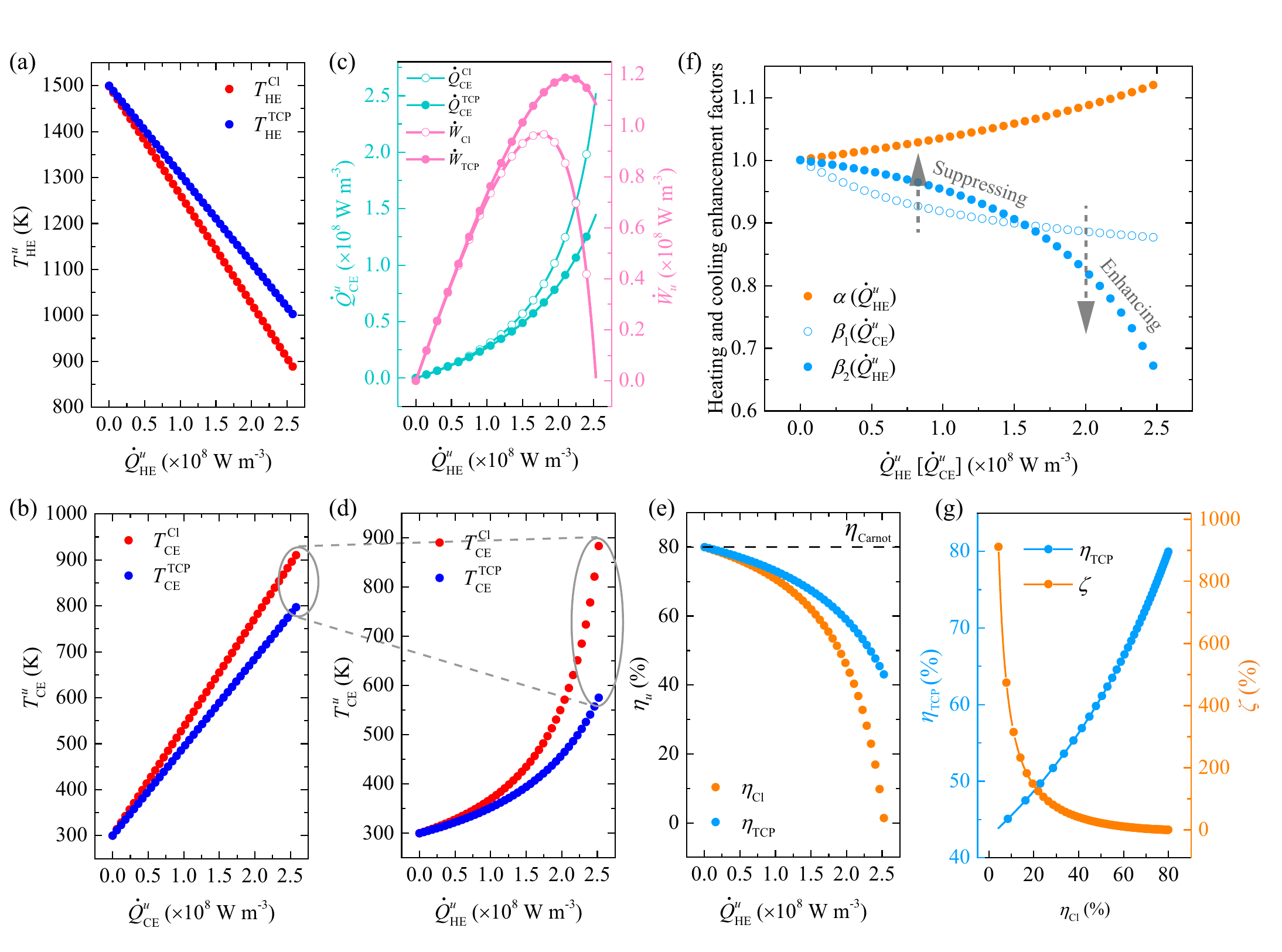}
\caption{\label{Fig. 5} Stirling engine parameters under specific operating conditions.}
 \end{figure}

Furthermore, Fig. 6(c) demonstrates that the engine power output $\dot W_u$ ($u=\mathrm{Cl},\mathrm{TCP}$) initially increases and then decreases with increasing $\dot Q_\mathrm{HE}^u$. This relationship can be attributed to two competing factors. Firstly, according to the principle of energy conservation, increasing $\dot Q_\mathrm{HE}^u$ enhances $\dot W_u$ (factor 1). However, the input and output heat flux affect the temperatures of the heating and cooling fluids, respectively. As $\dot Q_\mathrm{HE}^u$ [$\dot Q_\mathrm{CE}^u$] increases, the hot end temperature $T_\mathrm{HE}^u$ [$T_\mathrm{CE}^u$] decreases [increases], thereby reducing the thermal efficiency $\eta_u$ and subsequently decreasing the engine power output $\dot W_u$ (factor 2). The combined effect of these factors is reflected in the $W_u-\dot Q_\mathrm{HE}^u$ diagram. Similar trends have been observed in related research studies \cite{AKSOY2013,ALMAJRI2017}, where the engine power first increases and then decreases with increasing engine speed. The decreasing trend was attributed to insufficient heat transfer due to a shorter heat transfer time \cite{AKSOY2013}. The Stirling engine model developed in this study provides another mechanism: when the temperatures of finite waste heat and cold sources are influenced by the input and output heat, the aforementioned phenomenon may occur even if the heat transfer is sufficient, due to the decreasing temperature difference between the heat and cold sources. This indicates that the Stirling engine model successfully incorporates the influence of limited waste heat and cold resources on thermal efficiency.

\subsection{Coupling enhancement of CMTC}

The coupling of the heating and cooling correlations in the presence of the CMTC effect on the heater and cooler is crucial for a comprehensive analysis of the heat engine system. In order to achieve this, the independent variable $\dot Q^u_\mathrm{CE}$ in the cooling correlations needs to be converted to $\dot Q^u_\mathrm{HE}$ using the irreversible condition of the heat engine system as described in Eq. (6). The results are shown in Fig. 6(d). By comparing Fig. 6(b) and Fig. 6(d), it can be observed that the CMTC effect can be significantly enhanced when $\dot Q^u_\mathrm{HE}$ or $\dot Q^u_\mathrm{CE}$ is large. This phenomenon is referred to as coupling enhancement.

To comprehensively understand this phenomenon, it is necessary to examine the trends between the heating (cooling) enhancement coefficients ($\alpha$, $\beta_1$, $\beta_2$) and $\dot Q_\mathrm{HE}$, as depicted in Fig. 6(f). It is important to note that there are two cooling enhancement coefficients presented, where $\beta_1$ represents the cooling enhancement effect before coupling, and $\beta_2$ represents the cooling enhancement effect after coupling. It is evident that the cooling enhancement effect before coupling initially exhibits suppression and subsequent enhancement compared to the effect after coupling. This phenomenon can be attributed to two competing factors. On one hand, the cooling enhancement effect is positively correlated with $\dot Q_\mathrm{CE}^u$. When $\dot Q_\mathrm{HE}^u=\dot Q_\mathrm{CE}^u$, since $\dot Q_\mathrm{HE}^{u'}$ corresponding to $\dot Q_\mathrm{CE}^u$ is always lower than $\dot Q_\mathrm{CE}^u$, $\beta_2(\dot Q_\mathrm{HE}^u)$ will show a tendency of suppression relative to $\beta_1(\dot Q_\mathrm{CE}^u)$. On the other hand, as depicted in Fig. 6(d), the cooling enhancement effect is significantly improved with an increase in $\dot Q_\mathrm{HE}^u$. Under the same $\dot Q_\mathrm{HE}^u$, there are two output heat flux densities: $\dot Q_\mathrm{CE}^\mathrm{TCP}$ and $\dot Q_\mathrm{CE}^\mathrm{Cl}$. The larger the $\dot Q_\mathrm{HE}^u$, the greater the difference between $\dot Q_\mathrm{CE}^\mathrm{TCP}$ and $\dot Q_\mathrm{CE}^\mathrm{Cl}$. This increased difference widens the gap between $T_\mathrm{CE}^\mathrm{TCP}$ and $T_\mathrm{CE}^\mathrm{Cl}$, ultimately leading to an enhancement in $\beta_2(\dot Q_\mathrm{HE}^u)$ relative to $\beta_1(\dot Q_\mathrm{CE}^u)$.

 \subsection{Ultrahigh Stirling engine efficiency improvement}

Coupling enhancement plays a crucial role in significantly improving the efficiency of the Stirling engine, especially when $\dot Q_\mathrm{HE}^u$ is very large, as illustrated in Fig. 6(e). To further examine the values of $\eta_\mathrm{TCP}$ for different $\eta_\mathrm{Cl}$, the extent of CMTC's contribution to the improvement in thermal efficiency from $\eta_\mathrm{Cl}$ to $\eta_\mathrm{TCP}$ was calculated and presented in Fig. 6(g). When combined with Fig. 6(c), it can be observed that as $\dot Q_\mathrm{HE}^u$ approaches $2.5\times 10^8\ \mathrm{W\ m^{-3}}$, $\dot W_\mathrm{Cl}$ tends towards zero, while $\dot W_\mathrm{TCP}$ remains at a relatively high value. This allows $\eta_\mathrm{TCP}$ to remain above 40\% even as $\eta_\mathrm{Cl}$ approaches zero, resulting in an increase in thermal efficiency of nearly 900\%.

To gain a deeper understanding of the improvement in Stirling engine efficiency through CMTC, a theoretical analysis of the convection heat transfer between the heating (cooling) fluid and the hot (cold) end of the Stirling engine needs to be conducted.

Following Newton's cooling law, the heat flow rate between the heating and cooling fluid and the working substance on the hot and cold ends of the Stirling engine can be described as
\begin {subequations}
\begin{align}
&Q_\mathrm{HE}^u={h_\mathrm{H}^u}A_\mathrm{Hs}\left(T_\mathrm{Hf}^u-T_\mathrm{Hs}^u\right),\\
&Q_\mathrm{CE}^u={h_\mathrm{C}^u}A_\mathrm{Cs}\left(T_\mathrm{Cs}^u-T_\mathrm{Cf}^u\right),
\end{align}
\end {subequations}
where $h_\mathrm{H}^u$ ($h_\mathrm{C}^u$) represents the total convective heat transfer coefficient between the heating (cooling) fluid and the outer surface of the Cl/TCP shell on the hot (cold) end, $A_\mathrm{Hs}$ ($A_\mathrm{Cs}$) is the heat transfer area of the hot (cold) end shell's outer surface, $T_\mathrm{Hf}^u$ ($T_\mathrm{Cf}^u$) is the average temperature of the heating (cooling) fluid, and $T_\mathrm{Hs}^u$ ($T_\mathrm{Cs}^u$) is the average temperature of the hot (cold) end shell's outer surface. It should be noted that $T_\mathrm{Hs}^u$ ($T_\mathrm{Cs}^u$) is positively correlated with $T_\mathrm{HE}^u$ ($T_\mathrm{CE}^u$).

As mentioned earlier, CMTC improves heat transfer performance by regulating the temperature of the heating (cooling) fluid in the flow channel. For the heating process in the heater, the temperatures of characteristic points in the flow channel with CMTC are consistently higher than those without CMTC, as depicted in Fig. 2(b1). The heat flow plot in Fig. 2(a1) demonstrates that CMTC significantly enhances the heat flow from the heating fluid outside the shell to the working substance inside the shell. It raises $T_\mathrm{HE}^\mathrm{Cl}$ to $T_\mathrm{HE}^\mathrm{TCP}$ by elevating $T_\mathrm{Hf}^\mathrm{Cl}$ to $T_\mathrm{Hf}^\mathrm{TCP}$.

The increased temperature difference between the hot and cold ends raises the Stirling engine efficiency from $\eta_\mathrm{Cl}$ to $\eta_\mathrm{TCP}$. Combining Eq. (16), Eq. (18), and Eq. (19), 
\begin{equation}
\begin{aligned}
 \zeta&=\frac{\frac{1}{T_\mathrm{HE}^\mathrm{Cl}}-\frac{T_\mathrm{CE}^\mathrm{TCP}}{T_\mathrm{CE}^\mathrm{Cl}T_\mathrm{HE}^\mathrm{TCP}}}{\frac{1}{T_\mathrm{CE}^\mathrm{Cl}}-\frac{1}{T_\mathrm{HE}^\mathrm{Cl}}}=\frac{1-\frac{T_\mathrm{HE}^\mathrm{Cl}}{T_\mathrm{HE}^\mathrm{TCP}}\beta_2}{\frac{T_\mathrm{HE}^\mathrm{Cl}}{T_\mathrm{CE}^\mathrm{Cl}}-1}=\frac{1-\frac{\beta_2}{\alpha}}{\frac{T_\mathrm{HE}^\mathrm{Cl}}{T_\mathrm{CE}^\mathrm{Cl}}-1}\\
 &=\frac{\frac{T_\mathrm{CE}^\mathrm{Cl}}{T_\mathrm{HE}^\mathrm{Cl}}\left(1-\frac{\beta_2}{\alpha}\right)}{1-\frac{T_\mathrm{CE}^\mathrm{Cl}}{T_\mathrm{HE}^\mathrm{Cl}}}=\frac{\frac{T_\mathrm{CE}^\mathrm{Cl}}{T_\mathrm{HE}^\mathrm{Cl}}\left(1-\frac{\beta_2}{\alpha}\right)}{\eta_\mathrm{Cl}}.
 \end {aligned}
 \end {equation}
Multiplying both sides of Eq. (24) by -1 and adding $\frac{\left(1-\frac{\beta_2}{\alpha}\right)}{\eta_\mathrm{Cl}}$
\begin {equation}
\zeta=\left(\frac{1}{\eta_\mathrm{Cl}}-1\right)\left(1-\frac{\beta_2}{\alpha}\right)=\mathscr{F}_{10}^u\left(\dot Q_\mathrm{HE}^u\right)>0.
\end {equation}
When $\dot Q_\mathrm{HE}^u$ is sufficiently large, the Stirling engine efficiency without CMTC ($\eta_\mathrm{Cl}$) rapidly approaches zero due to the rapid convergence of $T_\mathrm{HE}^\mathrm{Cl}$ and $T_\mathrm{CE}^\mathrm{Cl}$. At this point, if CMTC plays a significant role by ensuring $\alpha >> 1$ and $\beta_2 << 1$, $\zeta$ will tend towards infinity. As a result, an ultrahigh thermal efficiency improvement can be achieved.

\subsection{Effect of various operating conditions on Stirling engine parameters}
For practical applications, the influence of changeable operating conditions on the effectiveness of CMTC in improving Stirling engine efficiency was considered. Various factors such as inlet temperatures, inlet velocities, densities, specific heat capacities, and thermal conductivities were investigated. The results, presented in Figs. 7-12, provide insights into the Stirling engine's performance enhancement with CMTC. Subfigures (a) and (b) depict the engine's power density and thermal efficiency as functions of input heat flux, respectively, showcasing the direct impact of CMTC on the engine's performance. Subfigures (c) illustrate the magnitude of Stirling engine efficiency improvement across all ranges of $\eta_\mathrm{Cl}$. To quantitatively analyze the heating and cooling enhancement performance of CMTC, subfigures (d) and (e) show the heating (cooling) enhancement coefficient plotted against $\dot Q_\mathrm{HE}^u$ [$\dot Q_\mathrm{CE}^u$] and $\eta_\mathrm{Cl}$, respectively. Furthermore, the effect of CMTC, which primarily depends on $T_\mathrm{Hf}^u$ and $T_\mathrm{Cf}^u$, is demonstrated in subfigures (f), highlighting the relationship between $T_\mathrm{Hf}^u$ [$T_\mathrm{Cf}^u$] and $\dot Q_\mathrm{HE}^u$ [$\dot Q_\mathrm{CE}^u$] with and without CMTC. These findings contribute to a comprehensive understanding of the CMTC's impact on Stirling engine efficiency under different operating conditions.
\subsubsection{Effect of inlet temperatures}
Fig. 7 provides insights into the impact of inlet temperatures on Stirling engine performance. Analyzing Figs. 7(a-c) and 7(f), it becomes evident that the inlet temperature difference between the heating and cooling fluids ($T_\mathrm{Hf,in}$-$T_\mathrm{Cf,in}$) determines the average temperature difference of these fluids [$T_\mathrm{Hf}^u-T_\mathrm{Cf}^u$]. A larger inlet temperature difference results in a greater temperature difference at both ends of the Stirling engine [$\left(T_\mathrm{Hf}^\mathrm{Cl}-T_\mathrm{Cf}^\mathrm{Cl}\right)\rightarrow \left(T_\mathrm{Hf}^\mathrm{TCP}-T_\mathrm{Cf}^\mathrm{TCP}\right)$], thereby enhancing power density and thermal efficiency under the same operating conditions [$\eta_\mathrm{Cl}\rightarrow \eta_\mathrm{TCP}$ and $\dot W_\mathrm{Cl}\rightarrow\dot W_\mathrm{TCP}$, Fig. 7(a-c)]. Furthermore, the increased temperature difference at the inlet significantly enhances the coupling effect [Fig. 7(d)]. Fig. 7(e) illustrates that the heating and cooling enhancement factors experience significant growth with increasing inlet temperature differences. Particularly for low initial efficiencies ($\eta_\mathrm{Cl}$), these effects make a substantial contribution to improving thermal efficiency [$\zeta$, Fig. 7(c)]. These findings suggest that higher inlet temperatures unlock the significant potential of CMTC when the Stirling engine's efficiency is constrained by finite waste heat and cold resources.
\begin{figure}[h!]
\centering
\includegraphics[width=\textwidth]  {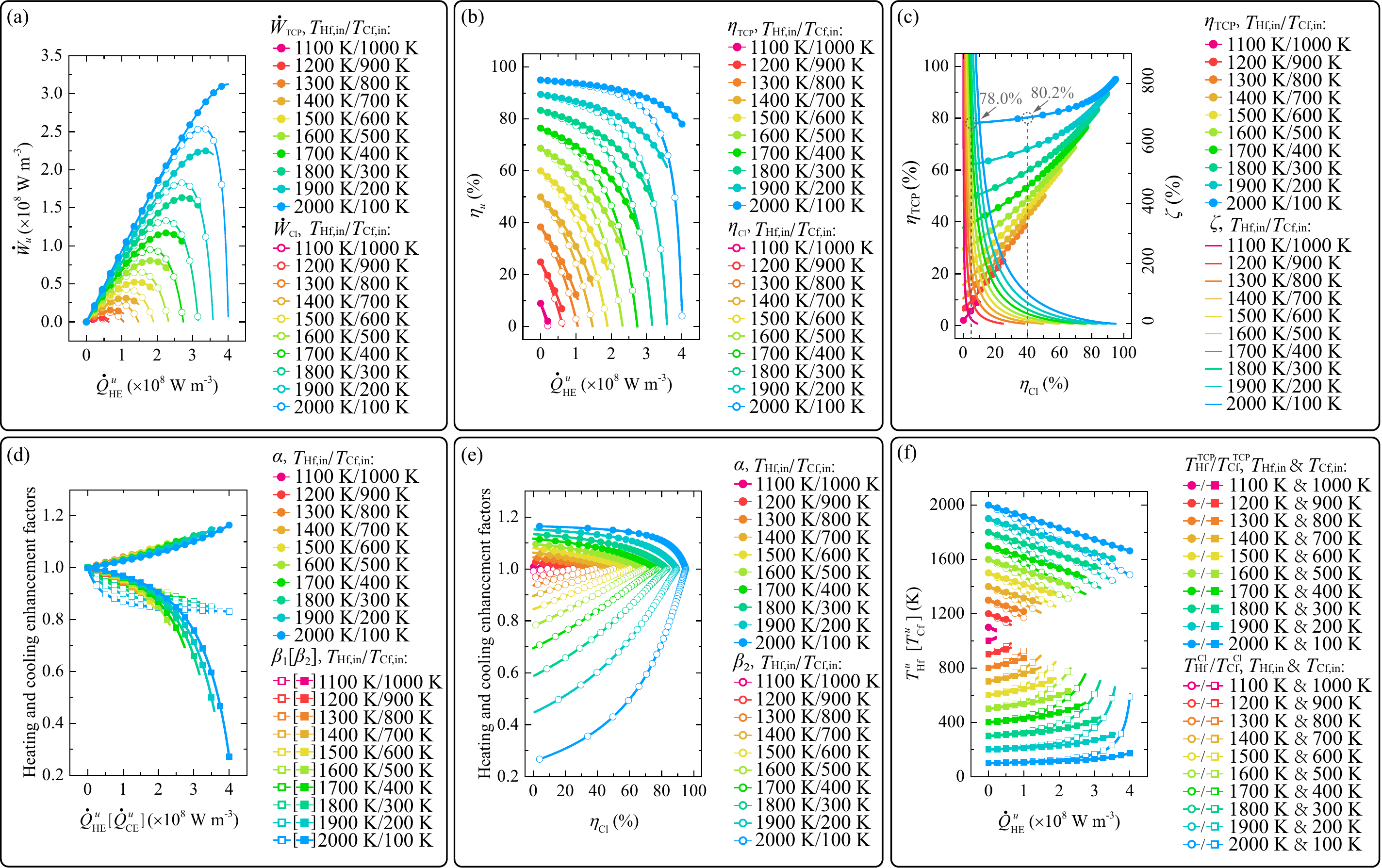}
\caption{\label{Fig. 5} Stirling engine parameters under different inlet temperatures.}
 \end{figure}

\subsubsection{Effect of inlet velocities}
Fig. 8 presents the influence of inlet fluid velocities on Stirling engine performance. Higher velocities of the heating and cooling fluids ($v_\mathrm{Hf,in}$, $v_\mathrm{Cf,in}$) result in more pronounced convective heat transfer. Consequently, the temperature difference at both ends of the Stirling engine increases, leading to improved power density and thermal efficiency [$\dot W_\mathrm{Cl}\rightarrow\dot W_\mathrm{TCP}$ and $\eta_\mathrm{Cl}\rightarrow \eta_\mathrm{TCP}$, Fig. 8(a,b)]. Additionally, higher $v_\mathrm{Hf,in}$ and $v_\mathrm{Cf,in}$ reduce the impact of $\dot Q_\mathrm{HE}^u$ and $\dot Q_\mathrm{CE}^u$ on the temperatures of the heating and cooling fluids ($T_\mathrm{Hf}^u$ and $T_\mathrm{Cf}^u$), bringing them closer to the fluid inlet temperatures [$T_\mathrm{Hf,in}$ and $T_\mathrm{Cf,in}$, Fig. 8(f)]. The effect of CMTC in enhancing Stirling engine efficiency becomes more prominent as $v_\mathrm{Hf,in}$ and $v_\mathrm{Cf,in}$ decrease [Fig. 8(c-e)]. This is because CMTC operates by adjusting $T_\mathrm{Hf}^u$ and $T_\mathrm{Cf}^u$. When $v_\mathrm{Hf,in}$ and $v_\mathrm{Cf,in}$ are relatively small, CMTC can significantly increase [decrease] $T_\mathrm{Hf}^u$ [$T_\mathrm{Cf}^u$], resulting in a larger temperature difference between the hot and cold ends ($T_\mathrm{Hf}^\mathrm{TCP}-T_\mathrm{Cf}^\mathrm{TCP}$, Fig. 8(f)]. Given that $v_\mathrm{Hf,in}$ and $v_\mathrm{Cf,in}$ might be limited by finite waste heat and cold resources, CMTC plays a crucial role in such scenarios.
\begin{figure}[h!]
\centering
\includegraphics[width=\textwidth]  {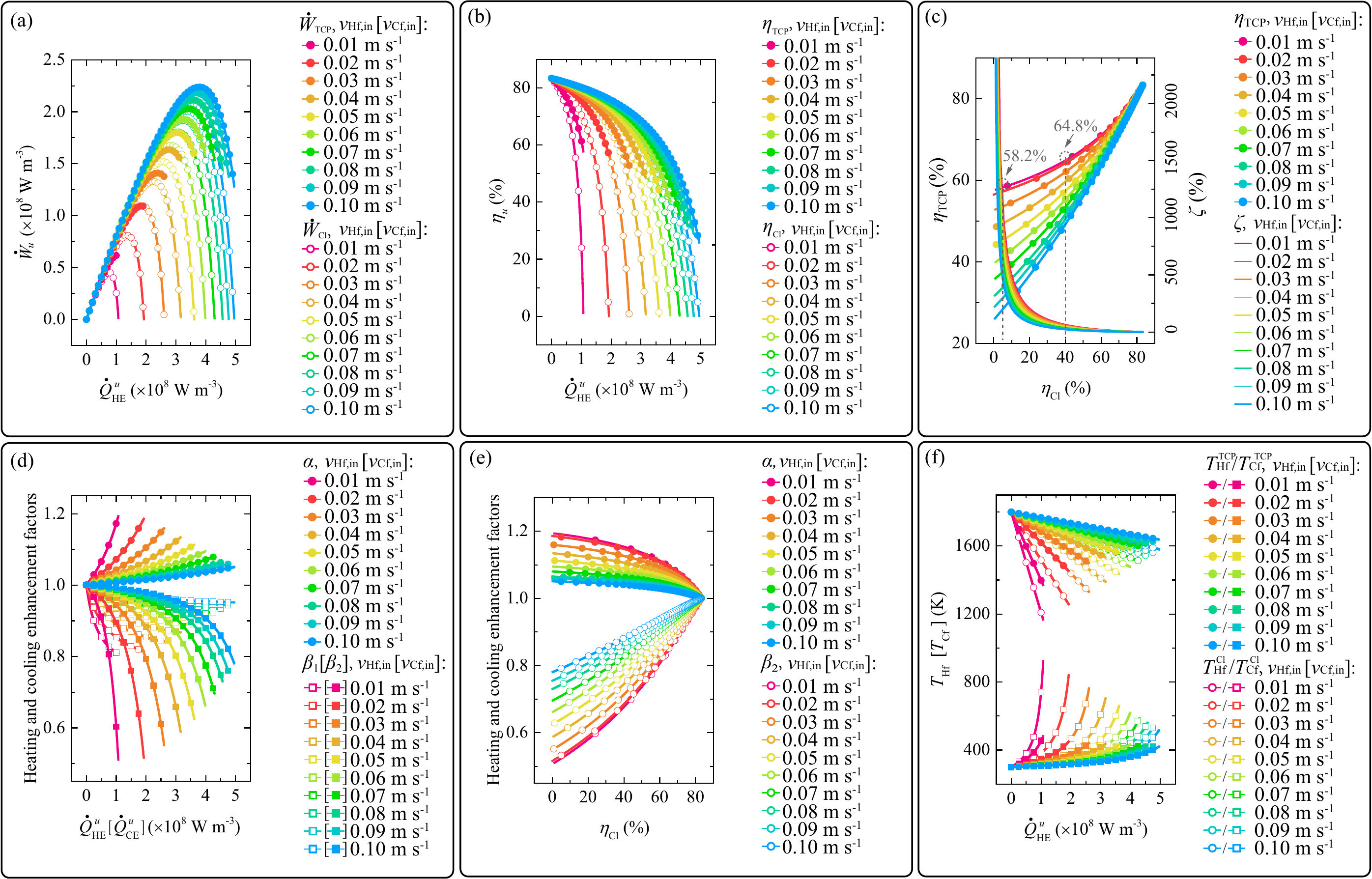}
\caption{\label{Fig. 5} Stirling engine parameters under different inlet velocities.}
 \end{figure}
\subsubsection{Effect of densities and specific heat capacities}
Figs. 9 and 10 demonstrate the impact of density and specific heat capacity of the heating and cooling fluids on Stirling engine performance. Increasing the density ($\rho_\mathrm{Hf}$, $\rho_\mathrm{Cf}$) and specific heat capacity ($c_\mathrm{Hf}$, $c_\mathrm{Cf}$) reduces the susceptibility of the heating and cooling fluid temperatures to the influence of $\dot Q_\mathrm{HE}^u$ and $\dot Q_\mathrm{CE}^u$ [Figs. 9(f) and 10(f)]. Consequently, enhancements in power density and thermal efficiency of the Stirling engine are observed, as depicted in Figs. 9(a,b) and 10(a,b). Similar to the decreasing $v_\mathrm{Hf,in}$ and $v_\mathrm{Cf,in}$, by reducing $\rho_\mathrm{Hf}$ [$\rho_\mathrm{Cf}$] and $c_\mathrm{Hf}$ [$c_\mathrm{Cf}$], $T_\mathrm{Hf}^u$ and $T_\mathrm{Cf}^u$ deviate significantly from $T_\mathrm{Hf,in}$ and $T_\mathrm{Cf,in}$ [Figs. 9(f) and 10(f)]. As a result, the effect of CMTC in enhancing Stirling engine efficiency is further amplified [Figs. 9(c) and 10(c)]. These findings suggest that if the density and specific heat capacity of the heating and cooling fluids are relatively low in practical applications, CMTC exhibits considerable potential for improving Stirling engine efficiency.
\begin{figure}[h!]
\centering
\includegraphics[width=\textwidth]  {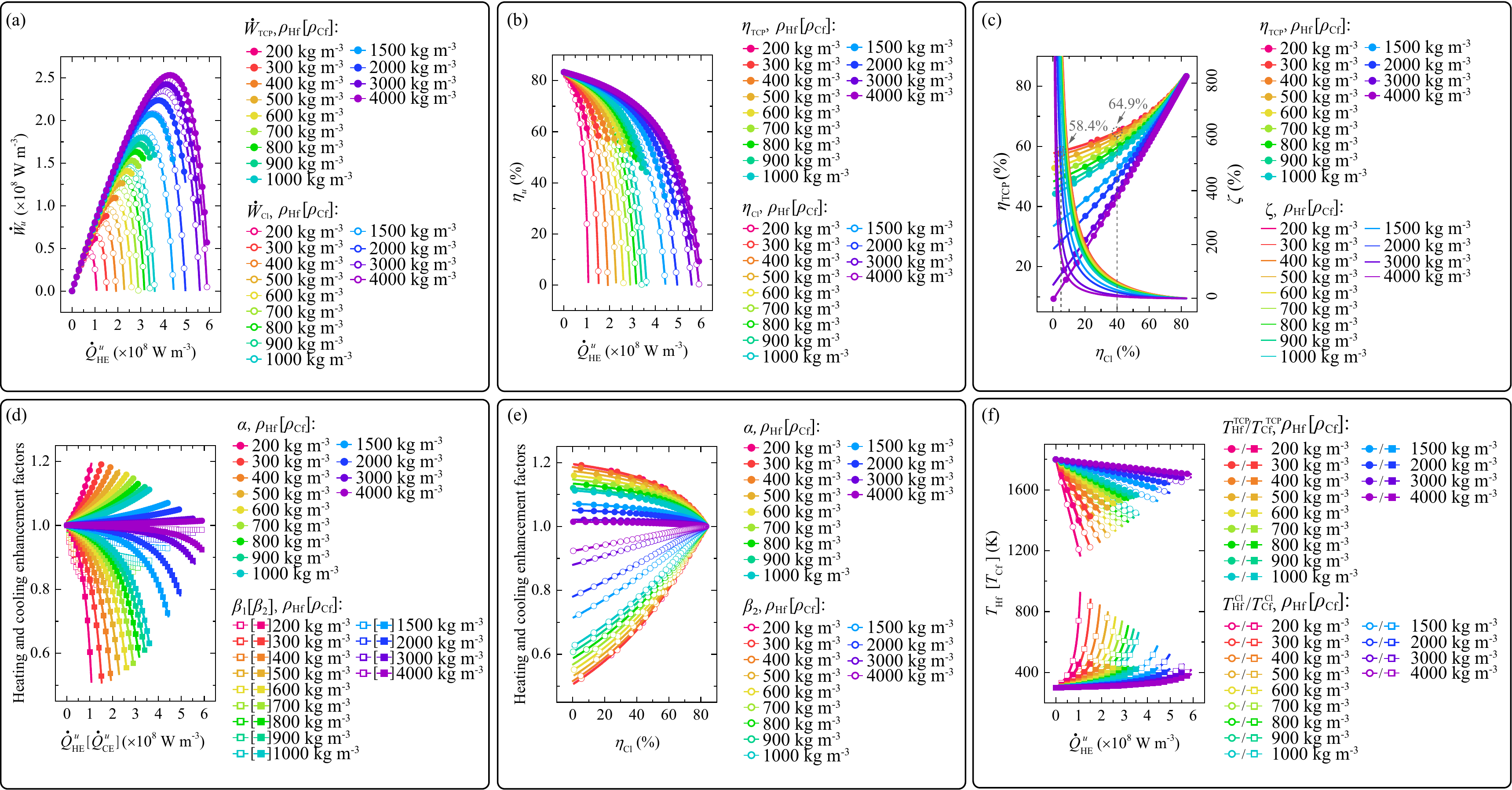}
\caption{\label{Fig. 5} Stirling engine parameters under different densities.}
 \end{figure}
 \begin{figure}[h!]
\centering
\includegraphics[width=\textwidth]  {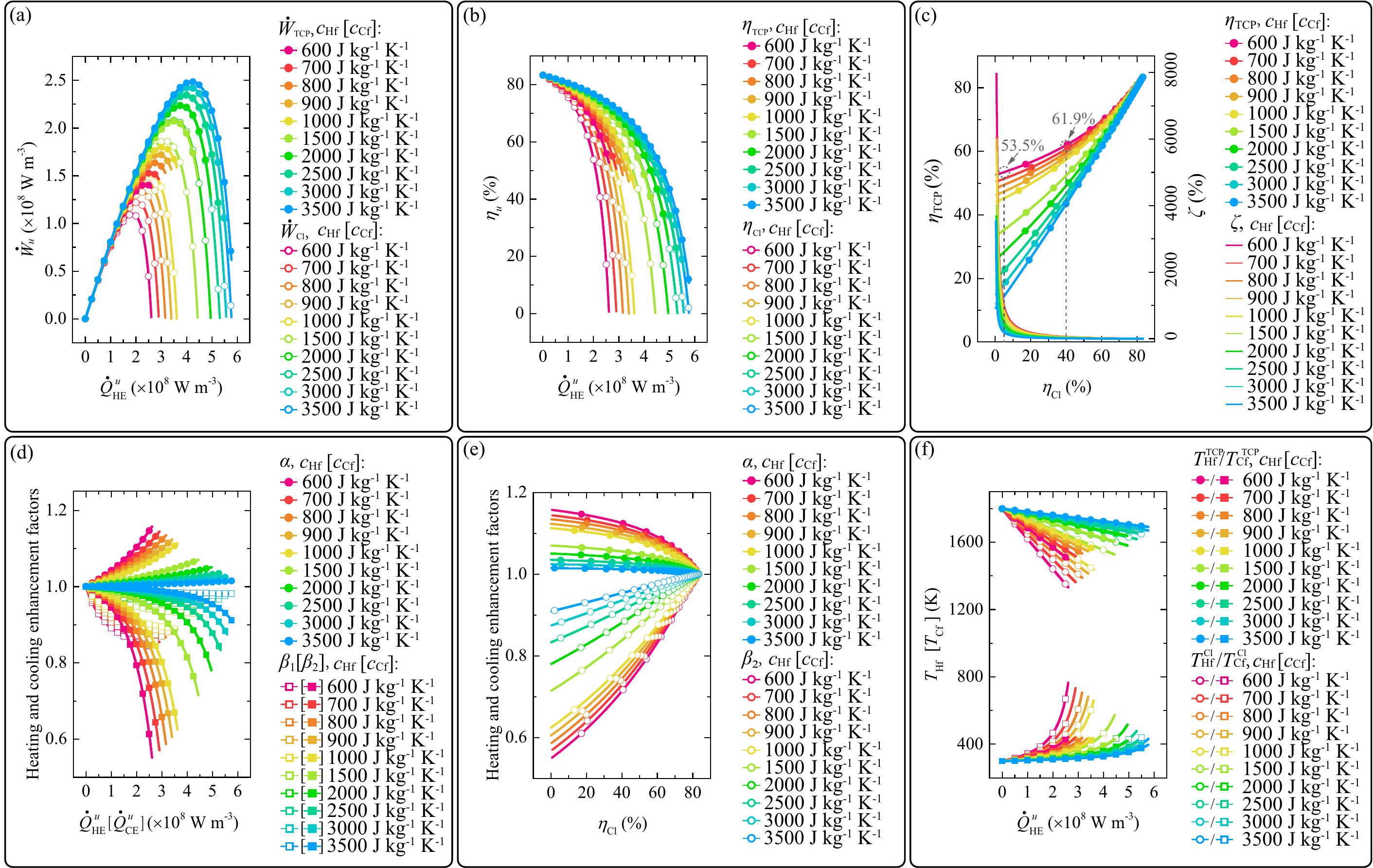}
\caption{\label{Fig. 5} Stirling engine parameters under different specific heat capacities.}
 \end{figure}
\subsubsection{Effect of thermal conductivities}
Figs. 11 and 12 provide insight into the influence of thermal conductivities ($\kappa_\mathrm{Hf}$, $\kappa_\mathrm{Cf}$) of the heating and cooling fluids on Stirling engine performance. In the Stirling model, thermal conductivity exhibits two competing effects. Firstly, increased thermal conductivities enhance convective heat transfer, leading to a larger temperature difference ($T_\mathrm{HE}^u$-$T_\mathrm{CE}^u$) between the engine's ends and thereby improving thermal efficiency (referred to as Factor A). However, under limited waste heat or heat source resources, the heating and cooling fluid temperatures ($T_\mathrm{Hf}$, $T_\mathrm{Cf}$) are influenced by $\dot Q_\mathrm{HE}^u$ and $\dot Q_\mathrm{CE}^u$. Consequently, enhancing convective heat transfer through increased thermal conductivity, with a constant $\dot Q_\mathrm{HE}^u$, reduces the temperature difference ($T_\mathrm{HE}^u$-$T_\mathrm{CE}^u$), thus hindering thermal efficiency improvement (referred to as Factor B). The interplay between these factors yields a complex influence on Stirling engine efficiency.

\begin{figure}[h!]
\centering
\includegraphics[width=\textwidth]  {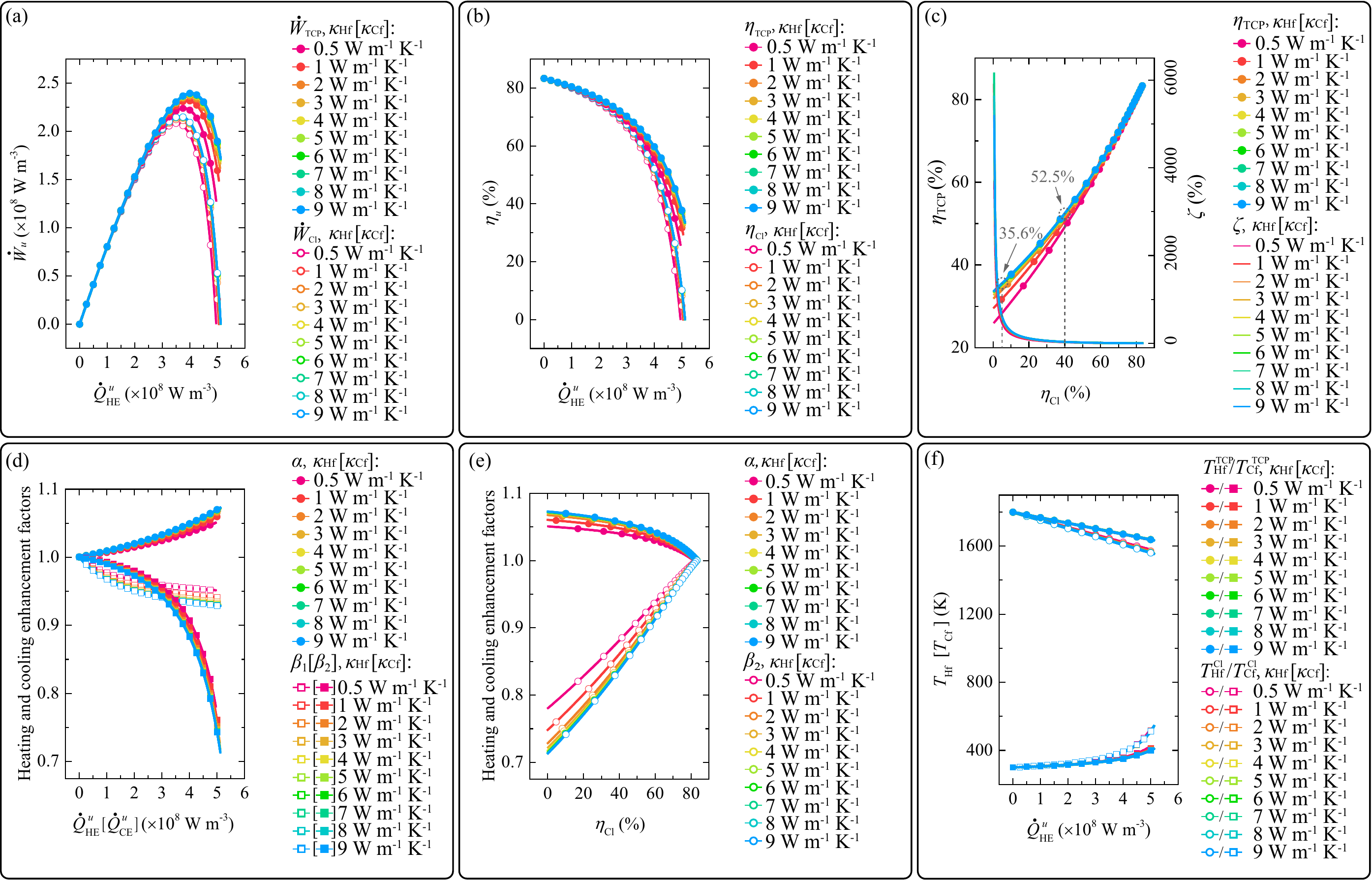}
\caption{\label{Fig. 5} Stirling engine parameters under different thermal conductivities with high densities and specific heat capacities.}
 \end{figure}
  \begin{figure}[h!]
\centering
\includegraphics[width=\textwidth]  {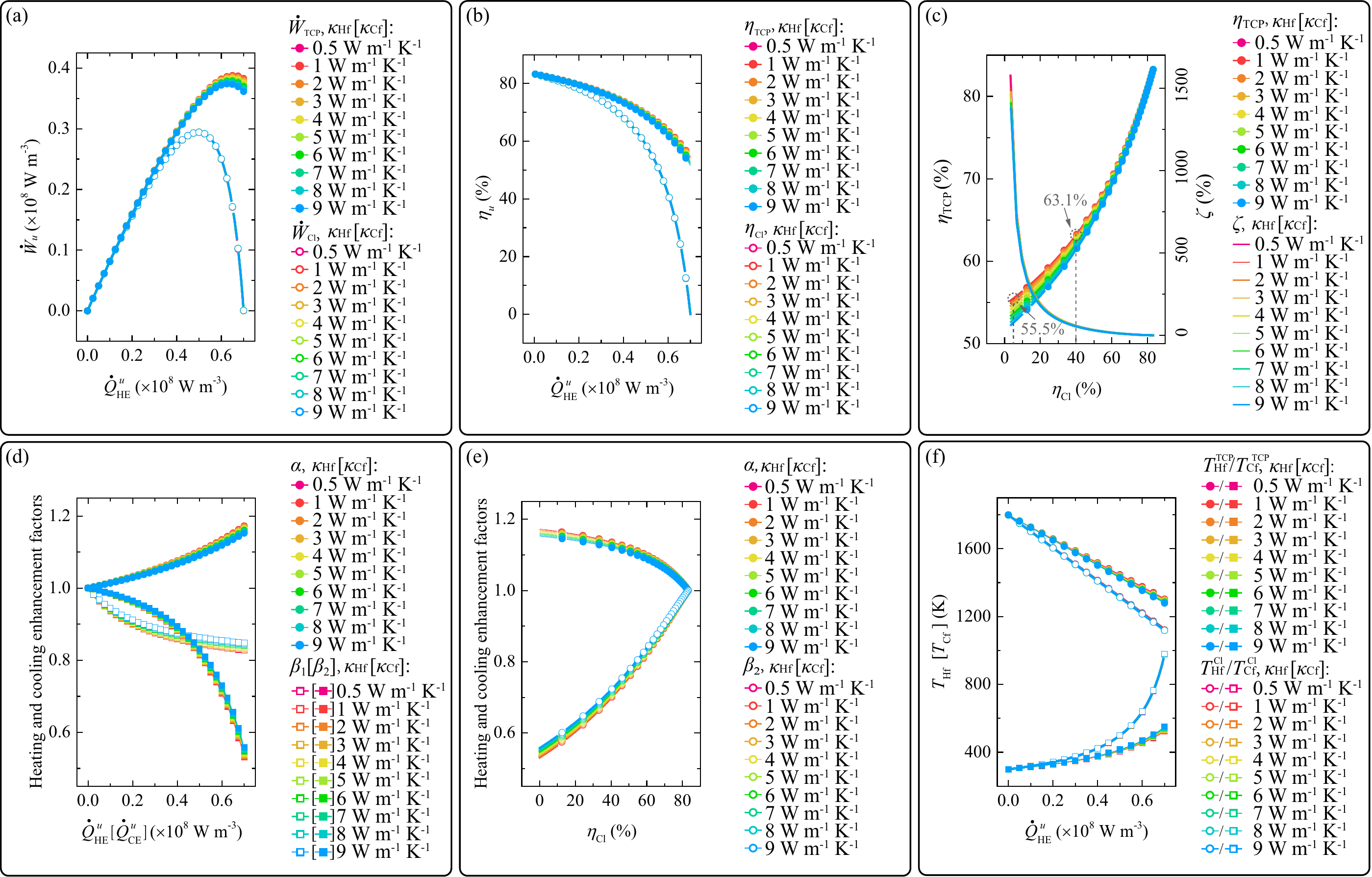}
\caption{\label{Fig. 5} Stirling engine parameters under different thermal conductivities with low densities and specific heat capacities.}
 \end{figure}
 
Further analysis highlights the significance of specific heat capacity ($c_\mathrm{Hf}$, $c_\mathrm{Cf}$) and density ($\rho_\mathrm{Hf}$, $\rho_\mathrm{Cf}$) of the heating and cooling fluids. When $c_\mathrm{Hf}$ [$c_\mathrm{Cf}$] and $\rho_\mathrm{Hf}$ [$\rho_\mathrm{Cf}$] are relatively large, Factor A dominates. In such cases, increasing $\kappa_\mathrm{Hf}$ and $\kappa_\mathrm{Cf}$ enhances power density ($\dot W_\mathrm{Cl}\rightarrow \dot W_\mathrm{TCP}$) and thermal efficiency ($\eta_\mathrm{Cl}\rightarrow \eta_\mathrm{TCP}$), as depicted in Fig. 11(a-c). Conversely, when $c_\mathrm{Hf}$ [$c_\mathrm{Cf}$] and $\rho_\mathrm{Hf}$ [$\rho_\mathrm{Cf}$] are relatively small, Factor B takes precedence. Elevating $\kappa_\mathrm{Hf}$ and $\kappa_\mathrm{Cf}$ diminishes power density ($\dot W_\mathrm{Cl}\rightarrow \dot W_\mathrm{TCP}$) and thermal efficiency ($\eta_\mathrm{Cl}\rightarrow \eta_\mathrm{TCP}$), as shown in Fig. 12(a-c). Furthermore, in such scenarios, the effect of CMTC on thermal efficiency enhancement intensifies with decreasing $\kappa_\mathrm{Hf}$ and $\kappa_\mathrm{Cf}$ [Fig. 12(c)]. The reduced thermal diffusivity of the heating and cooling fluids, resulting from lower $\kappa_\mathrm{Hf}$ and $\kappa_\mathrm{Cf}$, further underscores the role of CMTC in regulating $T_\mathrm{Hf}^u$ and $T_\mathrm{Cf}^u$.

\subsection{Overall analysis}
After analyzing various operating conditions, the mechanism of CMTC on Stirling engine efficiency can be explored by considering three key factors: inlet temperature difference ($\Delta T_\mathrm{in}$), specific heat capacity ($\dot C_\varepsilon$), and thermal diffusivity ($\mathscr{A}_\varepsilon$). The meanings of these factors were introduced in Sec. 2.3. Taking $\eta_\mathrm{Cl}=5\%$ as a reference, Fig. 13 presents the corresponding $\eta_\mathrm{TCP}$ under different operating conditions. Higher values of $\eta_\mathrm{TCP}$ indicate a greater contribution of CMTC to the improvement of thermal efficiency. By analyzing the influence of these factors on Stirling engine efficiency, more detailed results can be obtained.

Fig. 13(a) illustrates the influence of $\Delta T_\mathrm{in}$ on $\eta_\mathrm{TCP}$. It indicates that $\eta_\mathrm{TCP}$ increases rapidly with an increasing $\Delta T_\mathrm{in}$. As discussed in Sec. 3.4.1, this phenomenon can be attributed to the enhancing coupling effect. Therefore, placing the Stirling engine between waste heat and cold resources with higher temperature differences will significantly benefit the effect of CMTC in improving Stirling engine efficiency. This finding is particularly relevant to the existing operations that utilize cold energy from LNG to increase the temperature difference between both ends of the Stirling engine \cite{HAN2019561}.
 
 \begin{figure}[h!]
\centering
\includegraphics[width=12cm]  {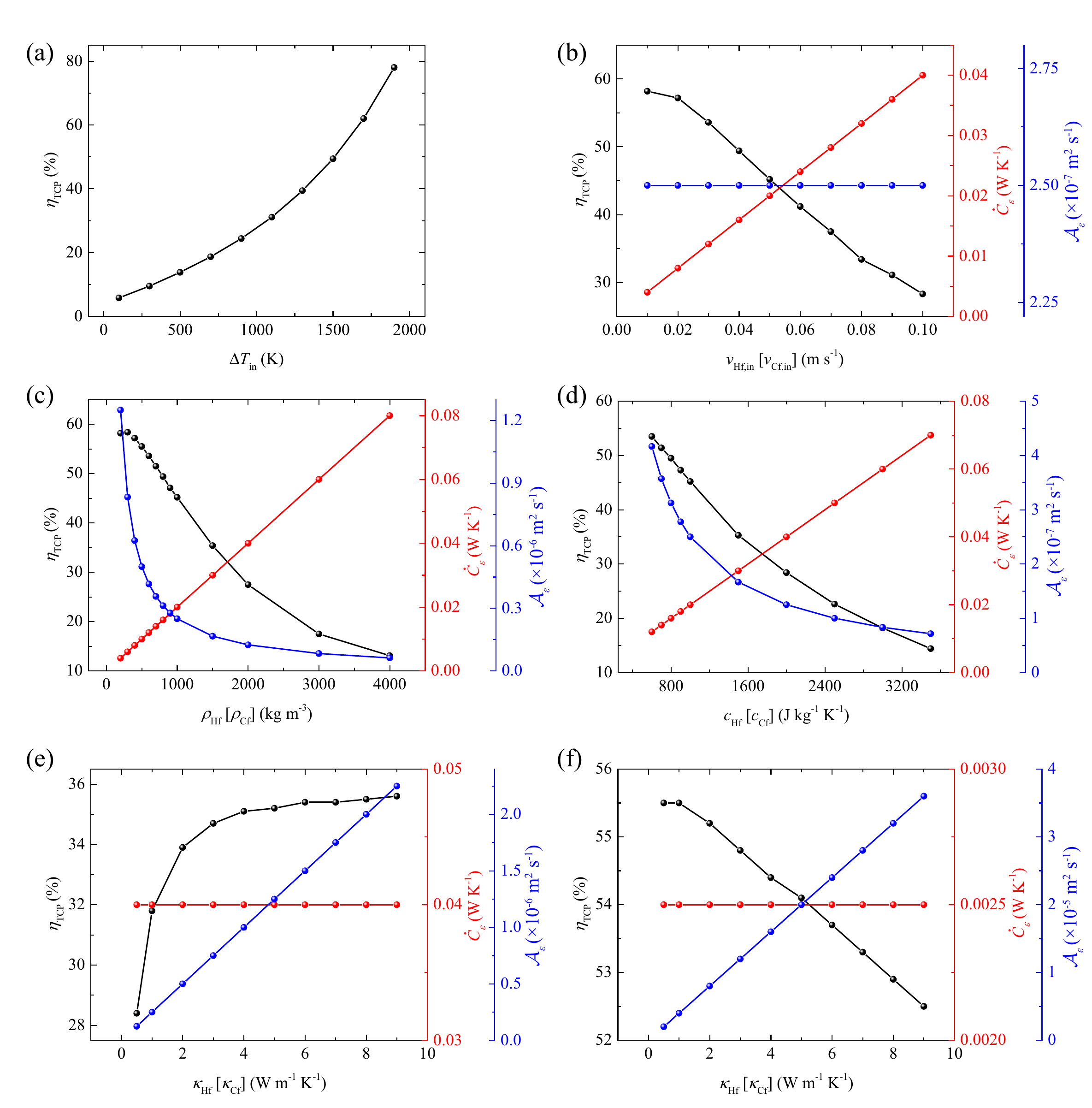}
\caption{\label{Fig. 5} Influence of $\Delta T_\mathrm{in}$, $\dot C_{\varepsilon}$, $\mathscr{A}_{\varepsilon}$ on $\eta_\mathrm{TCP}$ under different operating conditions.}
 \end{figure}
 
Next, Fig. 13(b) shows the influence of $\dot C_\varepsilon$ on $\eta_\mathrm{TCP}$ when $\mathscr{A}_\varepsilon$ is held constant. It can be observed that reducing $\dot C_\varepsilon$ improves $\eta_\mathrm{TCP}$. The reason behind this is that when $\dot C_\varepsilon$ is low, $T_\mathrm{Hf}^\mathrm{Cl}$ and $T_\mathrm{Cf}^\mathrm{Cl}$ are more susceptible to the influence of $\dot Q_\mathrm{HE}^u$. This provides a greater opportunity for CMTC to increase $T_\mathrm{Hf}^\mathrm{Cl}$ (or decrease $T_\mathrm{Cf}^\mathrm{Cl}$) towards $T_\mathrm{Hf}^\mathrm{TCP}$ (or $T_\mathrm{Cf}^\mathrm{TCP}$) in order to improve Stirling engine efficiency. To clarify this, the Stirling engine parameters were examined under different inlet velocities ($v_\mathrm{Hf,in}$ and $v_\mathrm{Cf,in}$), densities ($\rho_\mathrm{Hf}$ and $\rho_\mathrm{Cf}$), specific heat capacities ($c_\mathrm{Hf}$ and $c_\mathrm{Cf}$), and their corresponding $T_\mathrm{Hf}^u$ ($T_\mathrm{Cf}^u$) as shown in Figs. 8-10. These figures reveal that the Stirling engine operating under conditions of low $\dot C_\varepsilon$ (a combination of low $v_\varepsilon$, $c_\varepsilon$, and $\rho_\varepsilon$) can achieve significant thermal efficiency improvement through CMTC. Since $c_\varepsilon$ and $\rho_\varepsilon$ depend on the types of heat transfer fluids, this result is particularly relevant when waste heat and cold resources are limited, leading to low $v_\varepsilon$.

The effectiveness of CMTC in regulating temperatures $T_\mathrm{Hf}^\mathrm{Cl}$ and $T_\mathrm{Cf}^\mathrm{Cl}$ relies on the ability of the heating and cooling fluids to converge their temperatures, which is represented by the thermal diffusivity $\mathscr{A}_\varepsilon$. A low value of $\mathscr{A}_\varepsilon$ enables CMTC to effectively adjust the temperature distribution, making it highly suitable for such situations. However, it can be observed from Fig. 13(c,d) that reducing $\dot C_\varepsilon$ and increasing $\mathscr{A}_\varepsilon$ correspond to an increase in $\eta_\mathrm{TCP}$. This implies that increasing $\mathscr{A}_\varepsilon$ does not have an adverse effect on the CMTC mechanism. One possible explanation for this phenomenon is that the range of $10^{-7}\sim10^{-6}\ \mathrm{m^2\ s^{-1}}$ is relatively low for $\mathscr{A}_\varepsilon$, and variations within this range do not compromise the effectiveness of CMTC. This can be further verified by studying the influence of $\mathscr{A}_\varepsilon$ on CMTC while keeping $\dot C_\varepsilon$ constant, as shown in Fig. 13(e,f). In Fig. 13(e), where $\mathscr{A}_\varepsilon$ is around $10^{-6}\ \mathrm{m^2\ s^{-1}}$, increasing $\mathscr{A}_\varepsilon$ leads to a rise in $\eta_\mathrm{TCP}$. Note that the variation in $\mathscr{A}_\varepsilon$ in this case is caused by changes in $\kappa_\mathrm{Hf}$ and $\kappa_\mathrm{Cf}$. Therefore, the primary factor contributing to the increase in $\mathscr{A}_\varepsilon$ is the higher values of $\kappa_\mathrm{Hf}$ and $\kappa_\mathrm{Cf}$, resulting in enhanced convective heat transfer and improved Stirling engine efficiency. On the other hand, in Fig. 13(f), when $\mathscr{A}_\varepsilon$ is around $10^{-5}\ \mathrm{m^2\ s^{-1}}$, increasing $\mathscr{A}_\varepsilon$ leads to a decrease in $\eta_\mathrm{TCP}$. This indicates that when $\mathscr{A}_\varepsilon$ reaches a relatively high-value range, lower values of $\mathscr{A}_\varepsilon$ provide more potential for CMTC to improve Stirling engine efficiency, which is consistent with the above analysis.

To evaluate the effect of CMTC on the improvement of Stirling engine efficiency under different operating conditions, the existing efficiency improvement or range of the Stirling engine (based) system \cite{Sowale2019,UDEH2021,LAAZAAR2022,LAAZAAR2020,BULINSKI2019} was used as a reference. Then, the reference values for the Stirling engine efficiency without CMTC, denoted as $\eta_\mathrm{Cl}$, were set within the range of 5\% to 40\%. By analyzing the corresponding $\eta_\mathrm{TCP}$ values for various operating conditions, as shown in subfigures (c) of Figs. 7-12, the improvement effect could be quantitatively assessed. The results were organized and presented in Fig. 14, where the blue boxes represent the range of $\eta_\mathrm{TCP}$. It can be observed that the maximum magnitude of thermal efficiency improvement ($\zeta$) is 1460\%. In this particular scenario, the Stirling engine efficiency increased from 5\% to 78\%. Note that the initial low thermal efficiency of 5\% is primarily due to the limited availability of waste heat and cold resources. This remarkable improvement of the Stirling engine efficiency highlights the significant potential of CMTC, which greatly enhances energy utilization efficiency.

\begin{table}[htpb]\scriptsize
\centering
\tabcolsep=0.15cm
\caption{Thermal efficiency of various Stirling engine (based) models. }
\begin{tabular}{ccccc}
\hline
No. & Ref. & Methods & Key points&  \tabincell{c}{Thermal efficiency\\improvement/range} \\
\hline
1&\cite{Sowale2019}&   \tabincell{c}{Quasi-steady model\\+Genetic algorithm}& \tabincell{c}{Structural parameters\\+operating conditions}& 23\%$\rightarrow$27\%\\
2&\cite{UDEH2021}& \tabincell{c}{Non-ideal thermal model\\+MATLAB+Aspen plus®}& Stirling engine+ORC   &22.74\%$\rightarrow$37\%\\
3&\cite{LAAZAAR2022}&\tabincell{c}{Non-ideal adiabatic model\\+MATLAB}&\tabincell{c}{Structural parameters\\+operating conditions}&6.7\%-35.0\%\\
4&\cite{LAAZAAR2020}&\tabincell{c}{Non-ideal adiabatic model\\+MATLAB}&Operating conditions&8.0\%-41.5\%\\
5&\cite{BULINSKI2019}& \tabincell{c}{Non-equilibrium thermal model\\+ ANSYS
Fluent}&\tabincell{c}{Structural parameters\\+operating conditions}&5.4\%-25.7\%\\
6&\tabincell{c}{Present\\work}&\tabincell{c}{General Stirling engine model\\containing detailed heating\\and cooling processes\\+COMSOL Multiphysics}&\tabincell{c}{Structural optimization\\+operating conditions}&5\%$\rightarrow $78\% \\
\hline
\end{tabular}
\end{table}

 \begin{figure}[h!]
\centering
\includegraphics[width=10cm]  {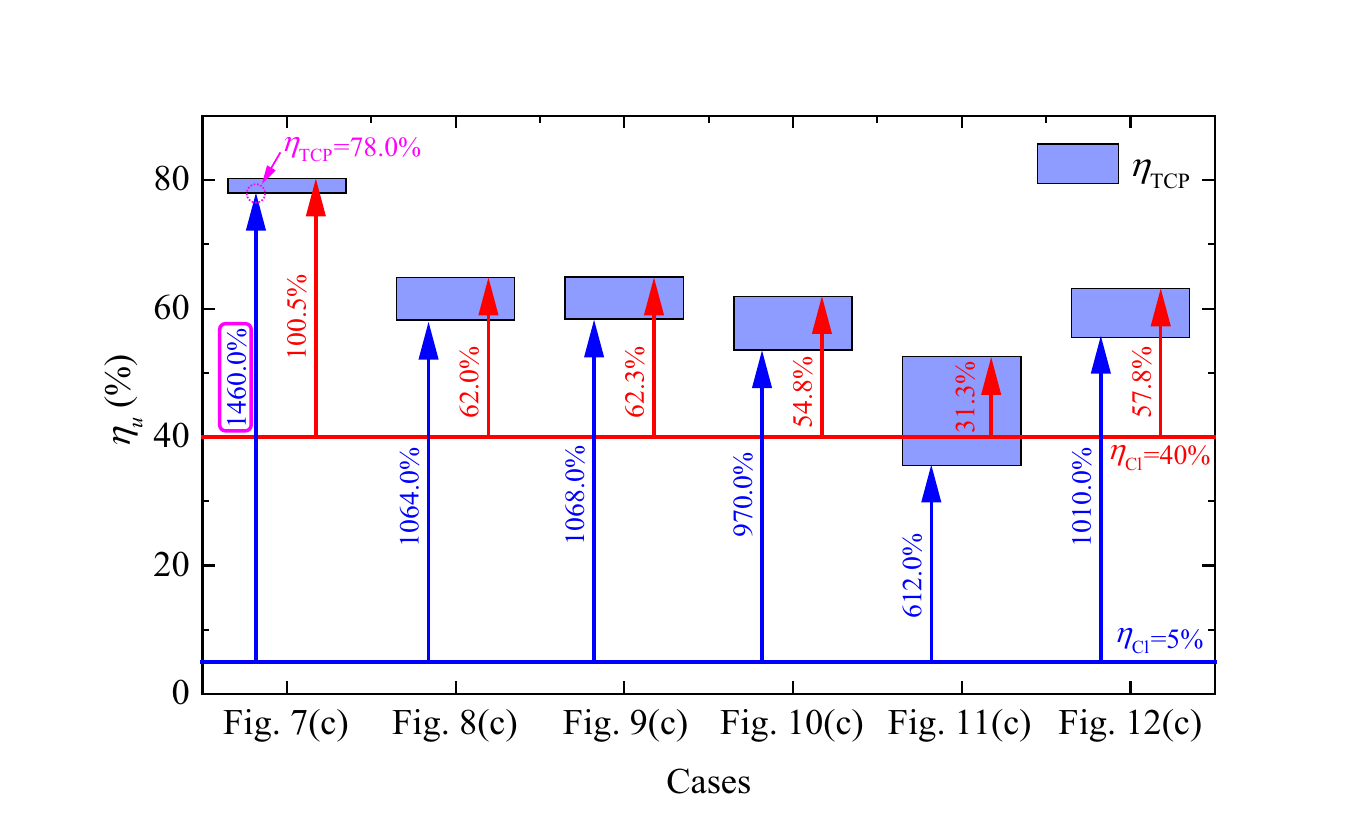}
\caption{\label{Fig. 5} Stirling engine efficiency improvement under different operating conditions.}
 \end{figure}

\subsection{Comparison with existing technologies and future applications}
The discussion above indicates that the Stirling engine with CMTC combines thermal metamaterials and the Stirling engine system. A comparison between the present work and existing technologies can be made as follows.

Regarding thermal metamaterials, a thermal concentrator has been successfully applied in a conduction heat transfer system, improving energy utilization efficiency \cite{Narayana,HTC2,SXY,HANTC2015}. The present work extends the application to a convection heat transfer system and combines it with a Stirling engine system. By utilizing equivalent anisotropic thermal conductivities, the temperature of the hot end can be passively increased, while the temperature of the cold end can be decreased. This coupling enhancement significantly increases the efficiency and output power of the Stirling engine. Therefore, this work expands the potential application of thermal metamaterials to heat engine systems.

In terms of Stirling engine systems, existing technologies for improving thermal efficiency and output power focus on structural optimization, enhancing thermal properties, and choosing operational conditions  \cite{MUNIR2020100664,LAI2016218,SONG2015,GHEITH2015, LAAZAAR2022}. The present work establishes a general framework for a Stirling engine system with waste heat and cold utilization, including detailed heating and cooling processes with temperature variation characteristics of heat-exchanging fluids. CMTC can be considered a simple structural optimization that increases the temperature difference on both ends of the Stirling engine without the need for enhancing thermal properties, enlarging heat transfer areas, or additional energy input. This approach is beneficial in terms of energy and material savings. The CMTC effect significantly improves thermal efficiency, especially when the engine efficiency is limited by finite waste resources. Therefore, this work offers a new approach to designing a Stirling engine system with waste heat and cold utilization.

For future applications, guidance can be taken from existing experimental operations of thermal concentrators under conduction heat transfer. For example, in Ref. \cite{SXY}, researchers used copper and expanded polystyrene as high and low thermal conductivity materials, respectively, and by alternately arranging these two materials, a thermal concentration function was achieved. Similar design approaches can be employed for CMTC with a comparable structure. Considering the liquid-solid heat transfer characteristics of the Stirling engine system with CMTC, it is necessary to protect the working substance inside the shells of the heater and cooler. The same anisotropic effect of CMTC can be achieved by drilling holes in high thermal conductivity solid shells, following the principles of the effective medium theory \cite{HJP, HUANG200687, TIAN2021121312, ZXCEPL2023}. Some fundamental theories in complex systems, graded materials, soft matter, nonlinear science, and nanotechnology can also be beneficial for practical design \cite{gao2007magnetophoresis,dong2004dielectric,huang2003dielectrophoresis,ye2008non,liu2013statistical,huang2005magneto,qiu2015nonstraight}. Additionally, various U-type tubes have been used in existing technologies to enhance the heat transfer performance of the heater and cooler by enlarging the heat transfer area \cite{SONG2015}. Incorporating the core idea of heat flow concentration in CMTC with such operational techniques could be advantageous for practical applications.

\section{Conclusions}
In summary, convective meta-thermal concentration (CMTC) was introduced into the Stirling engine system in this study. A general model was developed to analyze the detailed heating and cooling processes, with a focus on investigating the impact of CMTC on thermal efficiency improvement. The key findings can be summarized as follows.

(1) CMTC was found to significantly enhance the efficiency of Stirling engines by passively increasing the temperature differences between the hot and cold ends. The introduction of CMTC demonstrated a coupling effect that led to improved Stirling engine efficiency. Notably, CMTC exhibited a remarkable enhancement even when the efficiency approached zero due to the finite availability of waste heat and cold resources.

(2) The effectiveness of CMTC in improving Stirling engine efficiency was influenced by various operating conditions, including inlet temperatures, velocities, specific heat capacities, densities, and thermal conductivities. This effect was particularly prominent when the working fluids were more susceptible to changes in input/output heat flux within the Stirling engine. Three key factors, namely inlet temperature differences, heat capacity rates, and thermal diffusivities, were further examined. It was observed that when the fluids possessed high inlet temperature differences, low heat capacity rates, and low thermal diffusivities, CMTC had a greater potential for regulating the temperature distribution of the heat-exchanging fluids, thereby increasing the temperature difference across the Stirling engine and ultimately improving thermal efficiency. Comparative analysis between Stirling engines with and without CMTC under different operating conditions revealed a maximum magnitude of thermal efficiency improvement of 1460\%.

(3) The integration of CMTC into the Stirling engine system can be considered as a combination of thermal metamaterials and the Stirling engine itself. This integration extends the application of thermal concentrators to heat engine systems and offers new avenues for designing Stirling engines that effectively utilize waste heat and cold resources. The realization of CMTC was guided by experimental operations of thermal concentrators under conduction heat transfer systems, while the effective medium theory provided a valuable approach for designing a shell with anisotropic thermal conductivity, which is necessary for CMTC implementation.

From the above, CMTC represents a promising approach for designing Stirling engine systems with waste heat and cold utilization, as it enables ultrahigh thermal efficiency without requiring additional energy input, enhancements in thermal properties, or expansion of heat transfer areas. Furthermore, combining CMTC with existing optimization technologies holds potential for further practical applications. The findings of this study contribute to advancing the field of Stirling engine design and the efficient utilization of waste heat and cold resources.

\section*{Author contribution}
\textbf{Xinchen Zhou:} Conceptualization, Data curation, Formal analysis, Investigation, Methodology, Software, Validation, Visualization, Writing - original draft, review \& editing. \textbf{Xiang Xu:} Writing - review \& editing. \textbf{Xiaoping Ouyang:} Conceptualization. \textbf{Jiping Huang:} Conceptualization, Supervision, Funding acquisition, Resources, Project administration, Writing - review \& editing.

\section*{Declaration of Competing Interest}
The authors declare no competing interests.

\section*{Acknowledgments}
J.H. and X.Z. acknowledge the financial support provided by the National Natural Science Foundation of China under Grant No. 12035004, the Science and Technology Commission of Shanghai Municipality under Grant No. 20JC1414700, and the Innovation Program of Shanghai Municipal Education Commission under Grant No. 2023ZKZD06. X.X. acknowledges financial support from the National Key R$\&$D Program of China (2022YFC3005800).

\bibliography{mybib}
\bibliographystyle{elsarticle-num}
\end{document}